\pdfoutput=1

\documentclass[11pt,twoside,a4paper,cmspaper,final,collab]{cms-tdr}
\def\svnVersion{364528}\def\cmsCernNoTag{CERN-EP/2016-119}\def\cmsCernDate{\today}\def\cmsMessage{Published in the European Physical Journal C as \href{http://dx.doi.org/10.1140/epjc/s10052-016-4261-z}{\doi{10.1140/epjc/s10052-016-4261-z.}}}

\begin{document}\cmsNoteHeader{SUS-15-008}

\hyphenation{had-ron-i-za-tion}
\hyphenation{cal-or-i-me-ter}
\hyphenation{de-vices}
\RCS$Revision: 360918 $
\RCS$HeadURL: svn+ssh://svn.cern.ch/reps/tdr2/papers/SUS-15-008/trunk/SUS-15-008.tex $
\RCS$Id: SUS-15-008.tex 360918 2016-07-29 12:44:20Z gzevi $
\newcommand{\sslumi}{2.3\fbinv}
\newcommand{\MT}{\ensuremath{M_\mathrm{T}}\xspace}
\newcommand{\MTmin}{\ensuremath{M_\mathrm{T}^{\text{min}}}\xspace}
\newcommand{\Njets}{\ensuremath{N_\text{jet}}\xspace}
\newcommand{\Nbjets}{\ensuremath{N_{\PQb}}\xspace}
\newcommand{\WJ}{\ensuremath{\PW}+\text{jets}\xspace}
\newcommand{\WZ}{\ensuremath{\PW\PZ}\xspace}
\newcommand{\ttZ}{\ensuremath{\ttbar\PZ}\xspace}
\newcommand{\ttW}{\ensuremath{\ttbar\PW}\xspace}
\newcommand{\qqWW}{\ensuremath{\PW^{\pm}\PW^{\pm}}\xspace}
\newcommand{\susytop}{\ensuremath{{\PSQt_1}}\xspace}
\newcommand{\sbottom}{\ensuremath{{\PSQb_1}}\xspace}
\newcommand{\Totttt}{\ensuremath{\mathrm{T}1\mathrm{tttt}}\xspace}
\newcommand{\Tftttt}{\ensuremath{\mathrm{T}5\mathrm{tttt}}\xspace}
\newcommand{\Tfttcc}{\ensuremath{\mathrm{T}5\mathrm{ttcc}}\xspace}
\newcommand{\TfqqqqWW}{\ensuremath{\mathrm{T}5\mathrm{qqqq}\PW\PW}\xspace}
\newcommand{\TsttWW}{\ensuremath{\mathrm{T}6\mathrm{tt}\PW\PW}\xspace}
\newcommand{\TfttbbWW}{\ensuremath{\mathrm{T}5\mathrm{ttbb}\PW\PW}\xspace}
\newcommand{\ptRatio}{\ensuremath{\pt^\text{ratio}}\xspace}
\newcommand{\ptRel}{\ensuremath{\pt^\text{rel}}\xspace}
\newcommand{\miniIso}{\ensuremath{I_\text{mini}}\xspace}
\newcommand{\MGAMCNLO} {\textsc{MadGraph5\_aMC@NLO}\xspace}
\newcommand{\pb}{\ensuremath{\,\text{pb}}\xspace}
\newcommand{\fb}{\ensuremath{\,\text{fb}}\xspace}
\newcommand{\x}{\ensuremath{\phantom{0}}}
\ifthenelse{\boolean{cms@external}}{\providecommand{\cmsTable[1]}{\relax{#1}}}{\providecommand{\cmsTable[1]}{\resizebox{\textwidth}{!}{#1}}}

\cmsNoteHeader{SUS-15-008}
\title{Search for new physics in same-sign dilepton events in proton-proton collisions at $\sqrt{s} = 13\TeV$}

\date{\today}

\abstract{
A search for new physics is performed using events with two isolated same-sign leptons, two or more jets, and missing transverse momentum. The results are based on a sample of proton-proton collisions at a center-of-mass energy of 13\TeV recorded with the CMS detector at the LHC, corresponding to an integrated luminosity of 2.3\fbinv. Multiple search regions are defined by classifying events in terms of missing transverse momentum, the scalar sum of jet transverse momenta, the transverse mass associated with a $\PW$ boson candidate, the number of jets, the number of $\PQb$ quark jets, and the transverse momenta of the leptons in the event. The analysis is sensitive to a wide variety of possible signals beyond the standard model. No excess above the standard model background expectation is observed. Constraints are set on various supersymmetric models, with gluinos and bottom squarks excluded for masses up to 1300 and 680\GeV, respectively, at the 95\% confidence level. Upper limits on the cross sections for the production of two top quark-antiquark pairs (119\fb) and two same-sign top quarks (1.7\pb) are also obtained. Selection efficiencies and model independent limits are provided to allow further interpretations of the results.
}

\hypersetup{%
pdfauthor={CMS Collaboration},%
pdftitle={Search for new physics in same-sign dilepton events in proton-proton collisions at sqrt(s) = 13 TeV},%
pdfsubject={CMS},%
pdfkeywords={CMS, physics, SUSY}}

\maketitle

\section{Introduction}
\label{sec:intro}

Searches for new physics in final states with two leptons that have same-sign (SS) charges provide a powerful probe for searches of new physics,
both because standard model (SM) processes with this signature are few and have low cross sections,
and because this signature is produced in a large number of important new-physics scenarios.
Examples of the latter include the production of supersymmetric (SUSY) particles~\cite{Barnett:1993ea,Guchait:1994zk},
Majorana neutrinos~\cite{Almeida:1997em}, vector-like quarks~\cite{Contino:2008hi}, and SS top quark pairs~\cite{Bai:2008sk,Berger:2011ua}.
In the SUSY framework~\cite{Ramond:1971gb,Golfand:1971iw,Neveu:1971rx,Volkov:1972jx,Wess:1973kz,Wess:1974tw,Fayet:1974pd,Nilles:1983ge,Martin:1997ns},
the SS signature can arise through gluino pair production.
For example, the Majorana nature of the gluino allows gluino pairs to decay via SS charginos, yielding two SS W bosons.
Gluino pair production can also yield four W bosons, e.g., from the decay of four top quarks, which may result in the SS dilepton final state.
Alternatively, cascade decays of pair-produced squarks can lead to the SS dilepton signature.
Searches for new physics in the SS channel have been previously performed at the CERN LHC by the ATLAS~\cite{ATLAS:2012ai,Aad:2014pda,Aad:2016tuk}
and CMS~\cite{SUS-10-004,SUS-11-020,SUS-11-010,SUS-12-017,SUS-13-013} Collaborations.

This paper describes a search for new physics in the final state with two or more leptons and including a SS pair
($\PGmpm\PGmpm$, $\PGmpm\Pepm$, or $\Pepm\Pepm$, where $\PGm$ is a muon and $\Pe$ an electron).
The analysis is based on proton-proton (pp) collision data at $\sqrt{s} = 13$\TeV, corresponding to an integrated luminosity of \sslumi collected with the CMS detector in 2015.
The search strategy resembles that used in our analysis of 19.5\fbinv of data collected at $\sqrt{s}=8\TeV$~\cite{SUS-13-013},
which excluded gluino masses in the four top quark signature up to about 1050\GeV.
We design an inclusive analysis sensitive to a wide range of new-physics processes produced via strong interactions and yielding undetected particles in the final state.
The interpretations of the results consider $R$-parity conserving SUSY models~\cite{Farrar:1978xj}, as well as cross section limits on the production
of two top quark-antiquark (\ttbar) pairs and of two SS top quarks.
We also provide model independent limits to allow further interpretations of the results.
With respect to Ref.~\cite{SUS-13-013}, the kinematic regions are redefined and improvements in the event selection are implemented,
both of which increase the sensitivity to new-physics scenarios at $\sqrt{s}=13\TeV$.

\section{The CMS detector}

The central feature of the CMS apparatus is a superconducting solenoid of 6\unit{m} internal diameter, providing a magnetic field of 3.8\unit{T}.
Within the field volume are several particle detection systems.
Charged-particle trajectories are measured with silicon pixel and strip trackers, covering $0 \leq \phi < 2\pi$ in
azimuth and $\abs{\eta} < 2.5$ in pseudorapidity, where $\eta \equiv -\ln [\tan (\theta/2)]$ and $\theta$ is the
polar angle of the trajectory of the particle with respect to the counterclockwise beam direction.
The transverse momentum, namely the component of the momentum $p$ in the plane orthogonal to the beam, is defined as $\pt = p \sin \theta$.
Surrounding the silicon trackers, a lead tungstate crystal electromagnetic calorimeter and a brass and scintillator hadron calorimeter provide
energy measurements of electrons, photons, and hadronic jets in the range $\abs{\eta} < 3.0$.
Muons are identified and measured within $\abs{\eta} < 2.4$ by gas-ionization detectors embedded in the steel
flux-return yoke of the solenoid.
Forward calorimeters on each side of the interaction point encompass~$3.0 < \abs{\eta} < 5.0$.
The CMS trigger consists of a two-stage system. The first level of the CMS trigger system, composed of custom hardware processors,
uses information from the calorimeters and muon detectors to select events in a fixed time interval of less than 4\mus.
The high-level trigger (HLT) processor farm further decreases the event rate from around 100\unit{kHz} to less than 1\unit{kHz}, before data storage.
A more detailed description of the CMS detector can be found in Ref.~\cite{cmsDetector}.

\section{Event selection and Monte Carlo simulation}
\label{sec:eventsel}

Events are selected with two sets of HLT algorithms.
The first requires two very loosely isolated leptons, one satisfying $\pt > 17\GeV$
and the other satisfying $\pt > 8\GeV$ for a muon and 12\GeV for an electron.
The isolation is evaluated with respect to nearby tracks for a muon and to both tracks
and calorimetric objects for an electron.
The second set of triggers selects events with lowered \pt thresholds of 8\GeV and without a
restriction on the isolation, but requiring a hadronic activity $\HT^\text{HLT}>300\GeV$, where
$\HT^\text{HLT}$ is the scalar $\pt$ sum of all jets with $\pt > 40\GeV$
and $\abs{\eta} < 3.0$ identified by the HLT.
Typical trigger efficiencies for leptons satisfying the selection criteria described below
are 94\% (98\%) per muon (electron), with 100\% efficiency for the $\HT^\text{HLT}$ requirement.

In the subsequent analysis,
muon candidates are reconstructed by combining information from the silicon tracker and the
muon spectrometer in a global fit~\cite{Chatrchyan:2012xi}. A selection is
performed using the quality of the geometrical matching between the
tracker and muon system measurements.
We select muons with well-determined charge by imposing an additional criterion:
$\delta\pt(\mu)/\pt(\mu) < 0.2$, where $\delta\pt(\mu)$ is the uncertainty in the
measurement of the muon \pt from the global fit.

Electron candidates are reconstructed by combining clusters of energy
in the electromagnetic calorimeter with tracks in the silicon
tracker~\cite{Khachatryan:2015hwa}.
The identification is performed using a Boosted Decision Tree multivariate discriminant~\cite{Hocker:2007ht} based on shower shape and track
quality variables. The nominal selection criteria are designed to provide a maximum rejection of electron candidates
from multijet production while maintaining approximately 90\% efficiency for electrons
from the decay of $\PW$ or $\PZ$ bosons.
A relaxed selection on the multivariate discriminant is used to define ``loose'' criteria for electron identification.
To improve the accuracy of the electron charge reconstruction,
we require the position of the calorimeter deposit, relative
to the linear projection of the deposits in the pixel detector
to the inner calorimeter surface,
to be consistent with the charge determination from the full track fit.
Electrons originating from photon conversions are suppressed by rejecting candidates that are either without energy deposits
in the innermost layers of the tracking system, or that are associated with a displaced vertex compatible with a photon conversion.

Lepton candidates are required to be consistent with originating from the collision vertex
for which the summed $\pt^2$ of the associated physics objects is the largest.
The transverse (longitudinal) impact parameter of the leptons must not exceed 0.5 (1.0)\unit{mm} with respect to this vertex, and they must fulfill the requirement
$|d_\mathrm{3D}|/\sigma(d_\mathrm{3D}) < 4$, where $d_\mathrm{3D}$ is the three-dimensional impact parameter with respect to the vertex,
and $\sigma(d_\mathrm{3D})$ is its uncertainty from the track fit.

The charged leptons produced in decays of heavy particles, such as $\PW$ and $\PZ$ bosons
or SUSY particles (``prompt'' leptons), are typically spatially isolated from the hadronic activity in the event,
while leptons produced in hadron decays or in photon conversions, as well as hadrons misidentified as leptons, are usually
embedded in jets (``nonprompt'' leptons).
This distinction becomes less evident for systems with a high Lorentz boost, where decay products tend to overlap and jets may
contribute to the energy deposition around prompt leptons.
This problem is mitigated with an isolation definition constructed using the following three variables:

\begin{itemize}
\item the mini-isolation variable (\miniIso)~\cite{Rehermann:2010vq}, computed as the ratio of the scalar \pt sum
of charged hadrons, neutral hadrons, and photons within a cone of radius $\Delta R \equiv \sqrt{\smash[b]{(\Delta\eta)^2+(\Delta\phi)^2}}$ around the lepton
candidate direction at the vertex, to the transverse momentum of the lepton candidate ($\pt(\ell)$).
The cone radius $\Delta R$ depends on $\pt(\ell)$ as:
 \begin{equation}
\Delta R\left(\pt(\ell)\right) = \dfrac{10\GeV}{\min\left[\max\left(\pt(\ell), 50\GeV\right), 200\GeV\right]}.
  \end{equation}
The varying isolation cone definition takes into account the increased collimation of the decay products of a
hadron as its $\pt$ increases, and it reduces the inefficiency from accidental overlap between the lepton and jets
in a busy event environment.
The momentum estimate of each particle is performed by the particle-flow (PF) algorithm~\cite{CMS:2009nxa,CMS:2010byl},
which identifies individual particles through a combination of information from different detector components.
\item the ratio of the $\pt$ of the lepton to that of the closest jet within a distance $\Delta R = 0.4$:
 \begin{equation}
\ptRatio = \dfrac{\pt(\ell)}{\pt(\text{jet})},
  \end{equation}
where the definition of a jet is given below. In case of no jet within this distance, the value of \ptRatio is set to 1.
The \ptRatio variable is a measure of the isolation in a larger cone and improves the performance of the isolation definition,
especially for low-$\pt$ nonprompt leptons, which are more likely than high-$\pt$ leptons to appear in a jet that is wider than the \miniIso cone.
\item the \ptRel variable~\cite{Albajar:1986iu}, defined as the transverse momentum of the lepton relative to the residual momentum of the closest jet after lepton momentum subtraction:
  \begin{equation}
    \ptRel=\frac{\left|\left(\vec{p}(\text{jet})-\vec{p}(\ell)\right) \times \vec{p}(\ell)\right| }{|\vec{p}(\text{jet})-\vec{p}(\ell)|}.
  \end{equation}
This variable allows the identification of leptons that accidentally overlap with jets.
\end{itemize}

A lepton is considered to be isolated if the following condition is satisfied:
\begin{equation}
\label{eq:multiiso}
  \miniIso < I_1 \; \text{AND} \; ( \ptRatio > I_2 \; \text{OR} \; \ptRel > I_3 ).
\end{equation}
The values of $I_{i}$, with $i = 1,2,3,$ depend on the lepton flavor: because the probability to misidentify a lepton is higher for electrons,
tighter isolation values are used in this case (see Table~\ref{tab:isoWPs}).
In addition, a ``loose'' isolation criterion is defined as $\miniIso < 0.4$.

\begin{table}[h]
  \centering
    \topcaption{\label{tab:isoWPs} Values of the isolation parameters used in Eq.~(\ref{eq:multiiso}).}
    \begin{tabular}{lcc}
      \hline
      Isolation variable & Muons & Electrons  \\
      \hline
      $I_1$ & 0.16 & 0.12 \\
      $I_2$ & 0.76 & 0.80  \\
      $I_3\,(\GeVns)$ & 7.2 & 7.2 \\
      \hline
    \end{tabular}
\end{table}

Muons (electrons) are required to have $\pt > 10$ ($15$) $\GeV$ and $\abs{\eta} < 2.4$ ($2.5$);
at least one SS lepton pair with an invariant mass above $8\GeV$ must be present in the event.
In order to reduce backgrounds from inclusive production of the Z boson and from low-mass resonances decaying into lepton pairs,
the SS pair is rejected if there is an additional lepton in the event that satisfies loose requirements and that forms an opposite-sign, same-flavor pair with an invariant mass less than 12\GeV
or between 76 and 106\GeV with one of the two SS leptons.

Jets and missing transverse momentum ($\ETmiss$) are reconstructed with the PF algorithm.
We define \MET as the magnitude of the vector sum of all PF candidate transverse momenta~\cite{JME-13-003}.
For jet clustering, the anti-$k_t$ algorithm~\cite{Cacciari:2008gp} with a distance parameter of 0.4 is utilized.
Jets are required to satisfy quality requirements~\cite{Chatrchyan:2011ds} to remove those consistent with anomalous energy deposits.
After the estimated contribution from additional $\Pp\Pp$ interactions in the same or adjacent bunch crossings (pileup) is subtracted,
jet energies are corrected for residual nonuniformity and nonlinearity of the detector response
using simulation and data.
Jets are required to have \mbox{$\pt>40\GeV$} and to lie within the tracker acceptance \mbox{$\abs{\eta}<2.4$}.
Jets must be separated from loosely identified leptons by $\Delta R > 0.4$,
so that jets already employed for the calculation of lepton isolation variables are not considered further in the analysis.
We require $\Njets \geq 2$, where \Njets denotes the number of selected jets in the event.
The hadronic activity in the event ($\HT$) is defined as the scalar $\pt$ sum of the selected jets.

To identify jets originating from b quarks, the combined secondary vertex algorithm CSVv2~\cite{CMS-PAS-BTV-15-001} is used.
Jets with $\pt > 25\GeV$ and $\abs{\eta}<2.4$ are considered as b-tagged if they satisfy the requirements of the
medium working point of the algorithm.  These requirements result in approximately a 70\% efficiency for tagging a b quark jet, and
a less than 1\% mistagging rate for light-quark and gluon jets in \ttbar events. The number of b-tagged jets in the event is denoted as \Nbjets.

Monte Carlo (MC) simulation, which includes the contribution of pileup, is used to estimate the background
from SM processes with prompt SS leptons (see Section~\ref{sec:backgrounds})
and to calculate the efficiency for various new-physics scenarios.
The SM background samples are produced with the \MGAMCNLO~2.2.2 generator~\cite{MADGRAPH5}
at leading order (LO) or next-to-leading order (NLO) accuracy in perturbative quantum chromodynamics, with the exception of
diboson samples, which are produced with the \POWHEG~v2~\cite{Melia:2011tj,Nason:2013ydw} generator.
The NNPDF3.0LO~\cite{Ball:2014uwa} parton distribution functions (PDFs) are used for the simulated samples
generated at LO, and the NNPDF3.0NLO~\cite{Ball:2014uwa} PDFs for the samples generated at NLO.
Parton showering and hadronization are described using the \PYTHIA~8.205 generator~\cite{Sjostrand:2007gs}
with the CUETP8M1 tune~\cite{Skands:2014pea,CMS-PAS-GEN-14-001}.
The CMS detector response for the background samples is modeled with the \GEANTfour package~\cite{Geant}.
The signal samples are generated with \MGAMCNLO at LO precision, including up to two additional partons in the matrix element calculations;
parton showering and hadronization, as well as decays of SUSY particles, are simulated with \PYTHIA, while
the detector simulation is performed with the CMS fast simulation package~\cite{Abdullin:2011zz}.

\section{Search strategy}
\label{sec:strategy}

\begin{figure*}[t]
\centering
\includegraphics[width=.3\textwidth]{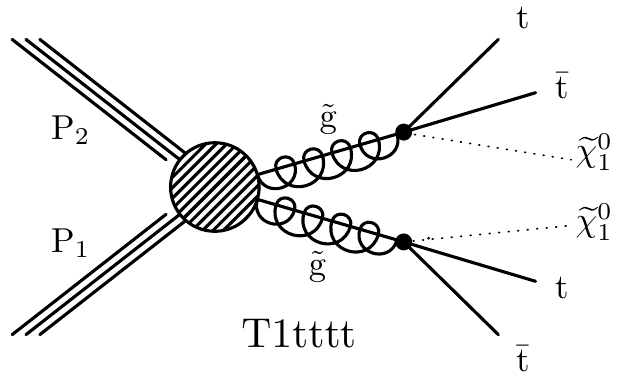}
\includegraphics[width=.3\textwidth]{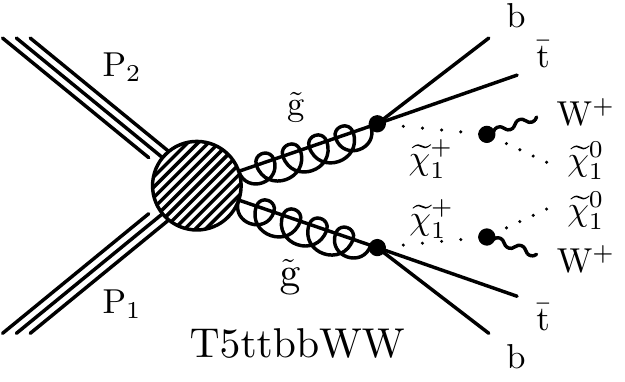}
\includegraphics[width=.3\textwidth]{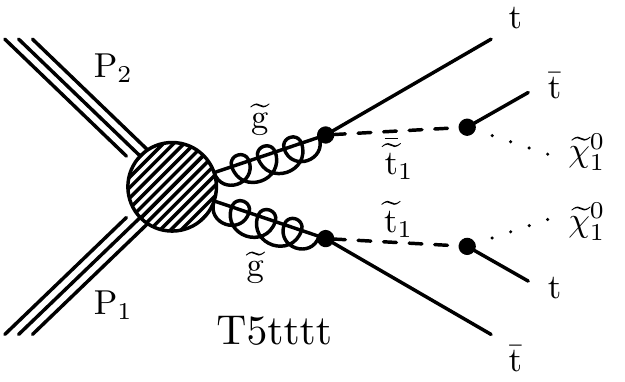}
\includegraphics[width=.3\textwidth]{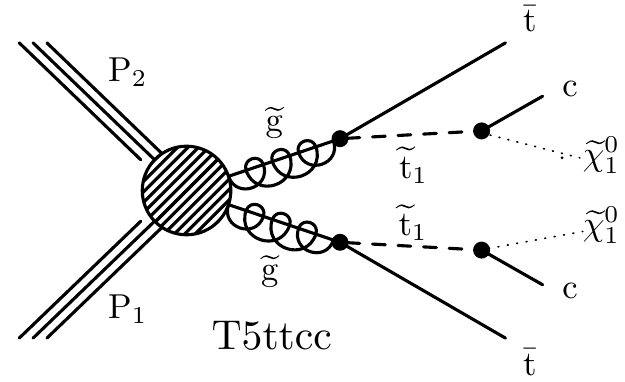}
\includegraphics[width=.3\textwidth]{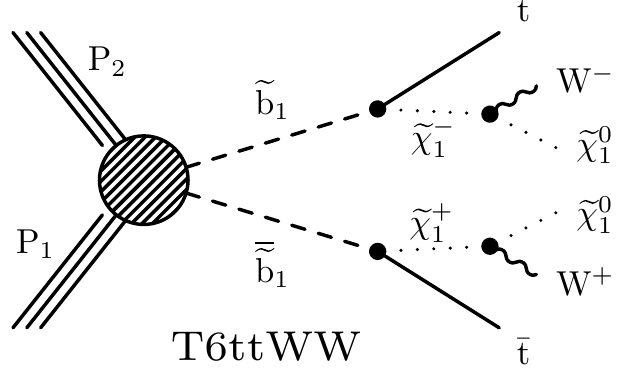}
\includegraphics[width=.3\textwidth]{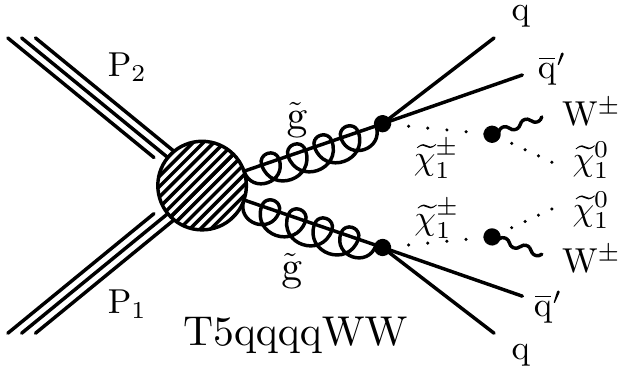}
\caption{Diagrams illustrating the simplified SUSY models used in this analysis.}
\label{fig:diagrams}
\end{figure*}

This analysis is designed as an inclusive search, sensitive to models matching two assumptions:
a strong-interaction production mechanism, leading to relatively large hadronic activity,
and the presence of undetected particles in the final state, yielding sizable \ETmiss.
In the process of defining the search strategy, $R$-parity conserving SUSY is taken as a guideline because of its rich variety of signatures.
In this context, signal models that can lead to the experimental signature of SS lepton pairs differentiate themselves in the
numbers of $\PW$ bosons, b jets, and light-flavor jets produced in the decays of SUSY particles.
In addition, the mass differences between the SUSY particles involved in the decay chains affect the energy spectra of the decay products,
resulting in differences between the models in the distributions of kinematic quantities such as the \pt of the leptons, \HT, and \ETmiss.

We consider SUSY scenarios in the context of simplified models of new-particle production~\cite{Alves:2011wf,Chatrchyan:2013sza}.
Models with four $\PW$ bosons and four b jets involve gluino pair production,
followed by the decay of each gluino through a chain containing third-generation squarks.
If the gluino is lighter than all squarks, and the top squark is the lightest squark,
the gluino undergoes a three-body decay mediated by an off-shell top squark.
If the dominant top squark decay is $\susytop \to \cPqt \PSGczDo$, where \PSGczDo is the lightest neutralino, taken to be
the stable, undetected, lightest SUSY particle (LSP),
then the gluino three-body decay is $\PSg \to \ttbar\PSGczDo$ (\Totttt model in Fig.~\ref{fig:diagrams}, upper left).
If instead the dominant top squark decay is $\susytop \to \PQb \PSGcpDo$,
the gluino three-body decay is $\PSg \to \cPaqt \PQb \PSGcpDo$ (\TfttbbWW model in Fig.~\ref{fig:diagrams}, upper middle);
the latter signature can also arise if the bottom squark is the lightest squark and decays as $\sbottom \to \cPqt \PSGcmDo$.
If the top squark is light enough to be on-shell and decays predominantly to a top quark and the LSP, gluinos decay through the chain
$\PSg \to \susytop\cPaqt \to \ttbar\PSGczDo$ (\Tftttt model in Fig.~\ref{fig:diagrams}, upper right).
If instead the top squark mainly decays to the charm quark and the LSP, gluinos decay as in the \Tfttcc model (Fig.~\ref{fig:diagrams}, lower left);
in this case only two $\PW$ bosons and two b jets are produced.

Events with four $\PW$ bosons and two b jets can arise
from bottom squark pair production, where each bottom squark decays to a top quark and a chargino,
and the chargino decays to an LSP and a (possibly off-shell) $\PW$ boson (\TsttWW model in Fig.~\ref{fig:diagrams}, lower middle).

Finally, SS lepton pairs can be produced in association with
large values of \HT, \ETmiss, and \Njets, but without b jets. In particular, events
with two $\PW$ bosons and four light-flavor quark jets can arise from gluino
pair production if each gluino decays to two light quarks and a chargino.
The two charginos can have the same charge and each decay to a $\PW$ boson
and the LSP (\TfqqqqWW model in Fig.~\ref{fig:diagrams}, lower right).
In the case that the difference in mass between the chargino and the LSP is small,
the $\PW$ bosons are off-shell and produce soft leptons.

To increase the sensitivity to new-physics scenarios, we categorize events based on their kinematic properties as follows.
First, three exclusive lepton selections are defined:
\begin{itemize}
\item high-high (HH) selection: two SS leptons, each with $\pt \geq 25\GeV$;
\item high-low (HL) selection: two SS leptons, one with $\pt \geq 25\GeV$ and the other with $10 \leq \pt < 25\GeV$;
\item low-low (LL) selection: two SS leptons, each with $10 \leq \pt < 25\GeV$.
\end{itemize}
The high lepton \pt threshold suppresses the contribution from nonprompt leptons;
hence the SM background in the HH region arises primarily from events with prompt SS leptons.
The nonprompt lepton background is largely contained in the HL region, where the high-\pt lepton is typically prompt and the low-\pt lepton nonprompt.
The LL region is characterized by a very small background since all processes where at least one lepton originates from an on-shell vector boson are
suppressed by the low-\pt requirements, while events with two nonprompt leptons are suppressed by the kinematic requirements described below;
the main residual contribution in this region is from nonprompt leptons.

Second, search regions (SR) are introduced so that the analysis is sensitive to a variety of new-physics scenarios.
SRs are defined separately for the HH, HL, and LL selections using the \HT, \MET, \Njets, and \Nbjets variables:
\Njets and \Nbjets separate signal from background for scenarios with a large production of jets and/or b jets, while \HT and \MET
increase sensitivity to models with different masses of SUSY particles.
In addition, we make use of the \MTmin variable, defined as:
\begin{equation}
\MTmin = \min\left[\MT(\ell_1,\ETmiss),\MT(\ell_2,\ETmiss)\right],
\end{equation}
where $\MT(\ell,\ETmiss) = \sqrt{\smash[b]{2\pt(\ell)\ETmiss(1-\cos\phi_{\ell,\ETmiss})}}$ is the transverse mass and $\phi_{\ell,\ETmiss}$
is the azimuthal angle difference between the directions of the lepton and of the missing transverse momentum~\cite{Arnison:1983rp}.
In the case of an SS lepton pair from \ttbar or \WJ processes, where one lepton is prompt and the other nonprompt, this variable has a cutoff near the $\PW$ boson mass;
consequently, the nonprompt lepton background is suppressed for SRs requiring $\MTmin > 120\GeV$ and is large for $\MTmin < 120\GeV$.
In order to better characterize the background we use a fine SR binning in kinematic regions where SM processes are abundant (e.g., low \MTmin and low \MET), while,
due to the low background, we use a coarser binning in regions with tight selections.

Finally, inclusive search regions in the HH and HL categories are defined in the tails of the \ETmiss and \HT variables;
the boundaries $\ETmiss>300\GeV$ and $\HT>1125\GeV$ (for $\ETmiss \leq 300\GeV$) are chosen so that each of these regions typically contains 1 background event.

A summary of the selection criteria is presented in Tables~\ref{tab:SRDefHH}--\ref{tab:SRDefLL}.
All SRs are non-overlapping. They are combined statistically to obtain the final results (Section~\ref{sec:results}).

\begin{table*}[h]
  \centering
    \topcaption{\label{tab:SRDefHH} SR definitions for the HH selection. The notation $^\mathrm{(*)}$ indicates that, in order to avoid overlaps with SR31, an upper bound $\ETmiss < 300\GeV$ is used for regions with $\HT > 300\GeV$.}
\cmsTable
{
\begin{tabular}{c|c|c|c|c|c|c}
\hline
$\Nbjets$ & $\MTmin$ (\GeVns) &$\ETmiss$ (\GeVns) & $\Nbjets$ & $\HT < 300\GeV$ & $\HT\in[300, 1125]\GeV$ & $\HT > 1125\GeV$  \\
\hline
\multirow{8}{*}{0} &  \multirow{4}{*}{${<}120$} & \multirow{2}{*}{ 50--200} & 2--4 & SR1 & SR2 & \multirow{30}{*}{SR32}\\
\cline{4-6}
& & & ${\geq}5$ & \multirow{7}{*}{SR3} & SR4 & \\
\cline{3-4} \cline{6-6}
& &  \multirow{2}{*}{${>}200^\mathrm{(*)}$} &  2--4 & &  SR5 & \\
\cline{4-4} \cline{6-6}
& & & ${\geq}5$ & & SR6 &  \\
\cline{2-4} \cline{6-6}
& \multirow{4}{*}{${>}120$} & \multirow{2}{*}{ 50--200} & 2--4 & & SR7 & \\
\cline{4-4} \cline{6-6}
& & & ${\geq}5$ & & \multirow{3}{*}{SR8} & \\
\cline{3-4}
& &  \multirow{2}{*}{${>}200^\mathrm{(*)}$} &  2--4 & & &  \\
\cline{4-4}
& & & ${\geq}5$ & & &  \\
\cline{1-6}
\multirow{8}{*}{1} & \multirow{4}{*}{${<}120$} & \multirow{2}{*}{ 50--200} & 2--4 & SR9 & SR10 & \\
\cline{4-6}
& & & ${\geq}5$ & \multirow{7}{*}{SR11} & SR12 & \\
\cline{3-4} \cline{6-6}
& & \multirow{2}{*}{ ${>}200^\mathrm{(*)}$} &  2--4 & &  SR13 & \\
\cline{4-4} \cline{6-6}
& & & ${\geq}5$ & & SR14 & \\
\cline{2-4} \cline{6-6}
 & \multirow{4}{*}{${>}120$} & \multirow{2}{*}{ 50--200} & 2--4 & & SR15 & \\
\cline{4-4} \cline{6-6}
& & & ${\geq}5$ &  & \multirow{3}{*}{SR16} & \\
\cline{3-4}
& & \multirow{2}{*}{ ${>}200^\mathrm{(*)}$} &  2--4 & & & \\
\cline{4-4}
& & & ${\geq}5$ & & & \\
\cline{1-6}
\multirow{8}{*}{2} & \multirow{4}{*}{${<}120$} & \multirow{2}{*}{ 50--200} & 2--4 & SR17 & SR18 & \\
\cline{4-6}
& & & ${\geq}5$ & \multirow{7}{*}{SR19} & SR20 &  \\
\cline{3-4} \cline{6-6}
& &  \multirow{2}{*}{ ${>}200^\mathrm{(*)}$} &  2--4 & & SR21 & \\
\cline{4-4}  \cline{6-6}
& & & ${\geq}5$ & & SR22 & \\
\cline{2-4} \cline{6-6}
& \multirow{4}{*}{${>}120$}  & \multirow{2}{*}{ 50--200} & 2--4 & &  SR23 & \\
\cline{4-4} \cline{6-6}
& & & ${\geq}5$ & & \multirow{3}{*}{SR24} &  \\
\cline{3-4}
& &  \multirow{2}{*}{ ${>}200^\mathrm{(*)}$} &  2--4 & &  & \\
\cline{4-4}
& & & ${\geq}5$ & &  & \\
\cline{1-6}
\multirow{4}{*}{${\geq}3$} & \multirow{2}{*}{${<}120$}  & 50--200 & ${\geq}2$ & SR25 & SR26 & \\
\cline{3-6}
& & ${>}200^\mathrm{(*)}$ & ${\geq}2$ & SR27 & SR28 & \\
\cline{2-6}
&  \multirow{2}{*}{${>}120$}  & \multirow{2}{*}{${>}50^\mathrm{(*)}$} &  \multirow{2}{*}{${\geq}2$} & \multirow{2}{*}{SR29} & \multirow{2}{*}{SR30} & \\
& &  & &  & &  \\
\hline
Inclusive & Inclusive & ${>}300$ & ${\geq}2$ & ------ & \multicolumn{2}{c}{SR31}  \\
\hline
\end{tabular}}
\end{table*}
\begin{table*}[h!]
\centering
\topcaption{\label{tab:SRDefHL} SR definitions for the HL selection. The notation $^\mathrm{(*)}$ indicates that, in order to avoid overlaps with SR25, an upper bound $\ETmiss < 300\GeV$ is used for regions with $\HT > 300\GeV$.}
\cmsTable
{
\begin{tabular}{c|c|c|c|c|c|c}
\hline
$\Nbjets$ & $\MTmin$ (\GeVns) & $\ETmiss$ (\GeVns) & $\Nbjets$ & $\HT < 300\GeV$ &  $\HT\in[300, 1125]\GeV$ & $\HT > 1125\GeV$  \\
\hline
\multirow{4}{*}{0} &  \multirow{4}{*}{${<}120$} & \multirow{2}{*}{50--200} & 2--4 & SR1 & SR2 & \multirow{17}{*}{SR26} \\
\cline{4-6}
& & & ${\geq}5$ & \multirow{3}{*}{SR3} & SR4 & \\
\cline{3-4} \cline{6-6}
& & \multirow{2}{*}{${>}200^\mathrm{(*)}$} &  2--4 & &  SR5 & \\
\cline{4-4} \cline{6-6}
& & & ${\geq}5$ & & SR6 &  \\
\cline{1-6}
\multirow{4}{*}{1} & \multirow{4}{*}{${<}120$} & \multirow{2}{*}{ 50--200} & 2--4 & SR7 & SR8 & \\
\cline{4-6}
& & & ${\geq}5$ & \multirow{3}{*}{SR9} & SR10 & \\
\cline{3-4} \cline{6-6}
& & \multirow{2}{*}{ ${>}200^\mathrm{(*)}$} &  2--4 & &  SR11 & \\
\cline{4-4} \cline{6-6}
& & & ${\geq}5$ & & SR12 & \\
\cline{1-6}
\multirow{4}{*}{2} & \multirow{4}{*}{${<}120$} & \multirow{2}{*}{ 50--200} & 2--4 & SR13 &  SR14 & \\
\cline{4-6}
& & & ${\geq}5$ & \multirow{3}{*}{SR15} & SR16 &  \\
\cline{3-4} \cline{6-6}
& & \multirow{2}{*}{ ${>}200^\mathrm{(*)}$} &  2--4 & & SR17 & \\
\cline{4-4}  \cline{6-6}
& & & ${\geq}5$ & & SR18 & \\
\cline{1-6}
\multirow{2}{*}{${\geq}3$} & \multirow{2}{*}{${<}120$}  & 50--200 & ${\geq}2$ & SR19 & SR20 & \\
\cline{3-6}
& & ${>}200^\mathrm{(*)}$ & ${\geq}2$ & SR21 & SR22 &  \\
\cline{1-6}
Inclusive & ${>}120$ & ${>}50^\mathrm{(*)}$  & ${\geq}2$ & SR23 & SR24 & \\
\hline
Inclusive & Inclusive & ${>}300$ & ${\geq}2$ & ------ & \multicolumn{2}{c}{SR25}  \\
\hline
\end{tabular}}
\end{table*}
\begin{table*}[h!]
  \centering
    \topcaption{\label{tab:SRDefLL} SR definitions for the LL selection. All SRs in this category require $\Nbjets \geq 2$.}
\begin{tabular}{c|c|c|c|c}
\hline
$\Nbjets$ &  $\MTmin$ (\GeVns) & $\HT$ (\GeVns) & $\ETmiss\in[50, 200]\GeV$ &  $\ETmiss> 200\GeV$ \\
\hline
0 & \multirow{4}{*}{${<}120$} &\multirow{5}{*}{${>}300$} & SR1 & SR2 \\
\cline{1-1} \cline{4-5}
1 & & & SR3 & SR4 \\
\cline{1-1} \cline{4-5}
2 & & & SR5 & SR6 \\
\cline{1-1} \cline{4-5}
${\geq}3$ & & & \multicolumn{2}{c}{SR7} \\
\cline{1-2} \cline{4-5}
Inclusive & ${>}120$ & & \multicolumn{2}{c}{SR8} \\
\hline
\end{tabular}
\end{table*}

\section{Backgrounds}
\label{sec:backgrounds}

Backgrounds in the SS dilepton final state can be divided into three categories:
\begin{itemize}
\item \textbf{Nonprompt leptons:} Nonprompt leptons are leptons from heavy-flavor decays, hadrons misidentified as leptons,
  muons from light-meson decays in flight, or electrons from unidentified conversions of photons in jets.
  Depending on the signal region, this background is dominated by \ttbar and \WJ processes;
  it represents the largest background for regions with low \MTmin and low \HT.
\item \textbf{SM processes with SS dileptons:} Standard model processes that yield an SS lepton pair include
  multi-boson production (considering \PW, \PZ, \PH, and prompt $\gamma$), single boson production in association with a \ttbar pair, and double-parton scattering.
  The dominant sources are \WZ and \ttW production, which contribute primarily to SRs with zero and
  ${\geq}1$ b jets, respectively.
  \WZ events contribute to the background when the $\PZ$ boson decays leptonically and is off-shell, when one of the $\PZ$-boson decay leptons is not identified,
  or when the $\PZ$ boson decays to $\tau$ leptons that result in a semileptonic final state.
  SM processes with SS dileptons are the largest background in the signal regions defined by tight kinematic selections.
\item \textbf{Charge misidentification:} Charge misidentification arises from events with opposite-sign isolated leptons in which the charge of an electron
  is misidentified, mostly due to severe bremsstrahlung in the tracker material. Overall, this is a small background.
\end{itemize}

The nonprompt lepton background is estimated from data using the ``tight-to-loose'' ratio method,
which was employed in previous versions of the analysis~\cite{SUS-10-004,SUS-11-020,SUS-11-010,SUS-12-017,SUS-13-013} but has been improved for the current study.
It is based on a control sample of events (application region) where one lepton fails the nominal (tight) selection but passes the loose requirements,
defined by relaxing the isolation selection for muons, and both the isolation and identification requirements for electrons.
Events in this control region are reweighted by the factor ${\epsilon_\mathrm{TL}/(1-\epsilon_\mathrm{TL})}$,
where $\epsilon_\mathrm{TL}$ is the probability for a nonprompt lepton that satisfies the loose selection to also satisfy the tight selection~\cite{SUS-10-004}.
Its value is measured in a multijet-enriched data set (measurement region), using events from single-lepton triggers after applying a selection designed to suppress
electroweak processes (Drell--Yan and \WJ) and after subtracting their residual contribution; this selection requires only one lepton in the event, $\ETmiss<20\GeV$, and $\MT<20\GeV$.
The measurement is made as a function of the lepton \pt and $\eta$, separately for each lepton flavor ($\mu$ or \Pe) and trigger (with or without isolation).

The method assumes that $\epsilon_\mathrm{TL}$ has the same value in the measurement and application regions.
The main sources of discrepancy are identified as differences in the momentum spectrum and the flavor of the parton producing the nonprompt lepton.
These two effects are mitigated in the following way. First, $\epsilon_\mathrm{TL}$ is parameterized as a function of $\pt^\text{corr}$, defined as the lepton \pt plus the energy in the isolation cone
exceeding the isolation threshold value~--- this quantity is highly correlated with the mother parton \pt, and thus the parameterization is robust against mother parton \pt variations.
The second effect, i.e., flavor dependence, is relevant for electrons only: while nonprompt muons originate predominantly from heavy-flavor decays, nonprompt electrons receive sizable
contributions from misidentified hadrons and conversions.
The effect of variations in the flavor composition is suppressed by adjusting the loose electron identification criteria so that the numerical value of $\epsilon_\mathrm{TL}$
for electrons from light flavors matches that for electrons from heavy flavors.
The loose lepton selection is defined based on MC studies, but we verify that $\epsilon_\mathrm{TL}$ is not significantly different in data events with and without b jets.

As a cross-check of the prediction, an alternative $\epsilon_\mathrm{TL}$ measurement, similar to that described in Ref.~\cite{ATLAS:2014kca}, is performed in the dilepton
control region where one of the leptons fails the impact parameter requirement. The predictions from the two methods are found to be consistent, both in MC samples and in data.

The background from SM processes with a prompt SS lepton pair is evaluated from simulation, accounting for both theoretical and experimental uncertainties.
The \WZ background is normalized to data in a control region requiring at least two jets, no b jets, $\ETmiss>30\GeV$, and three leptons, where two of the leptons form a same-flavor,
opposite-sign pair with an invariant mass within 15\GeV of the Z boson mass; the measured normalization factor is found to be compatible with unity
within about one standard deviation.
The MC simulation of \WZ production is used to relate the number of expected WZ events in the signal regions to the WZ event yield in the control region.

Finally, the charge misidentification background is estimated by reweighting events with opposite-sign lepton pairs by the charge misidentification probability.
For electrons this probability is obtained from simulated \ttbar events and from ${\Pe^\pm\Pe^\pm}$ data in the Z mass window,
and it lies in the range 10$^{-5}$--10$^{-3}$ depending on the electron \pt and $\eta$.
Studies of simulated events indicate that the muon charge misidentification probability is negligible.

\section{Systematic uncertainties}
\label{sec:systematics}

Systematic uncertainties can affect both the overall normalization and the
relative population of signal and background processes.
A summary of their effects on the SR yields is given in Table~\ref{tab:systSummary}.

Experimental systematic uncertainties are mostly the consequence of
differing event selection efficiencies in data and simulation.
Lepton identification and trigger efficiencies are computed with the ``tag-and-probe'' technique~\cite{Chatrchyan:2012xi,Khachatryan:2015hwa}
with an uncertainty of 2 and 4\%, respectively.
For signal samples, additional uncertainties of 4--10\% are included to account for differences in the lepton efficiency between the fast and \GEANTfour-based simulations.
The jet energy scale uncertainty varies between 2 and 8\%, depending on the jet $\pt$ and $\eta$.
Its impact is assessed by shifting the energy of each jet and propagating the variation to all dependent kinematic quantities (\HT, \ETmiss, \Njets, \Nbjets, and \MTmin);
correlation effects due to the migration of events from one SR to another are taken into account.
These variations yield estimated uncertainties of 2--10\%.
A similar approach is used to estimate the uncertainties associated with the $\PQb$ tagging efficiencies for light-flavor and b quark jets~\cite{CMS-PAS-BTV-15-001},
which are parameterized as a function of $\pt$ and $\eta$ and are found to be of order 5\% for the highly populated SRs.
The uncertainty in the modeling  of pileup is 1--5\% depending on the SR. The uncertainty in the integrated luminosity is 2.7\%~\cite{CMS-PAS-LUM-15-001}.

The background sources estimated from simulation are subject to
theoretical uncertainties related to unknown higher-order effects and to uncertainties in the knowledge of the PDFs.
The former are estimated by simultaneously varying the renormalization and factorization scales
up and down by a factor of two. The effect on the overall cross section is
found to be 13\% for \ttW events and 11\% for \ttZ events, while the effect
on the acceptance for the various SRs amounts to 3--8\% depending on \HT.
The magnitude of the uncertainty related to the PDFs is obtained
using variations of the NNPDF3.0 set~\cite{Ball:2014uwa}. The overall
uncertainty is $\sim 4\%$ for the \ttW and \ttZ samples.
Theoretical uncertainties are also considered for the remaining minor
backgrounds estimated from simulation: a similar procedure is used for
the \qqWW process, leading to an overall uncertainty of 30\%, while a 50\% uncertainty is assigned
to processes with a prompt $\gamma$ and to the sum of the other rare processes.
For all backgrounds estimated from simulation we account for the statistical uncertainty of the MC samples.

The remaining sources of uncertainty are those related to the
methods that are used to estimate the nonprompt lepton, charge misidentification,
and \WZ backgrounds.
An overall normalization uncertainty of 30\% is assigned to the
nonprompt lepton background prediction. This uncertainty accounts for the performance
of the method on simulated data and for the differences in the prediction from the
two alternative procedures described in Section~\ref{sec:backgrounds}.
An additional uncertainty is associated with the subtraction procedure
to remove Drell--Yan and \WJ events from the measurement region;
the overall effect on the nonprompt lepton background yield is 1--20\%,
depending on the SR considered, and is larger for high-\pt leptons.
Finally, we account for the statistical uncertainty in the number of events observed in
the application region.

The background from charge misidentification is assigned a systematic uncertainty of 26\%,
which corresponds to the difference between the ${\Pe^\pm\Pe^\pm}$ event yield in the Z mass window in data and simulation.

The uncertainty in the \WZ background is measured to be 30\% in the control region.
It includes statistical uncertainties and systematic uncertainties due to non-\WZ background subtraction.
Using the same procedure as described above, uncertainties in the extrapolation from the control
to the signal regions are assessed from the propagation of the
uncertainty in the jet energy scale and in the $\PQb$ tagging efficiencies.

\begin{table*}[!hbtp]
\centering
    \topcaption{Summary of systematic uncertainties in the event yields in the SRs.
The upper group lists uncertainties related to experimental factors for all processes whose yield is estimated from simulation;
the middle group lists uncertainties in these yields related to the event simulation process itself.
The lower group lists uncertainties for background processes whose yield is estimated from data.}
\label{tab:systSummary}
\begin{tabular}{lc}
\hline
Source          & Typical uncertainty (\%) \\
\hline
Lepton selection &  2 \\
Trigger efficiency &  4 \\
Jet energy scale & 2--10 \\
$\PQb$ tagging & 5 \\
Pileup & 1--5 \\
Integrated luminosity & 2.7 \\
\hline
Scale variations (\ttZ and \ttW) & 11--13 \\
Parton distribution functions (\ttW and \ttZ) & 4 \\
\qqWW normalization & 30 \\
Other backgrounds & 50 \\
Monte Carlo statistical precision & 1--30 \\
\hline
Nonprompt leptons & 30--36 \\
Charge misidentification & 26 \\
\WZ normalization & 30 \\
\hline
\end{tabular}
\end{table*}

\section{Results}
\label{sec:results}

\begin{figure*}[!hbtp]
\centering
\includegraphics[width=.35\textwidth]{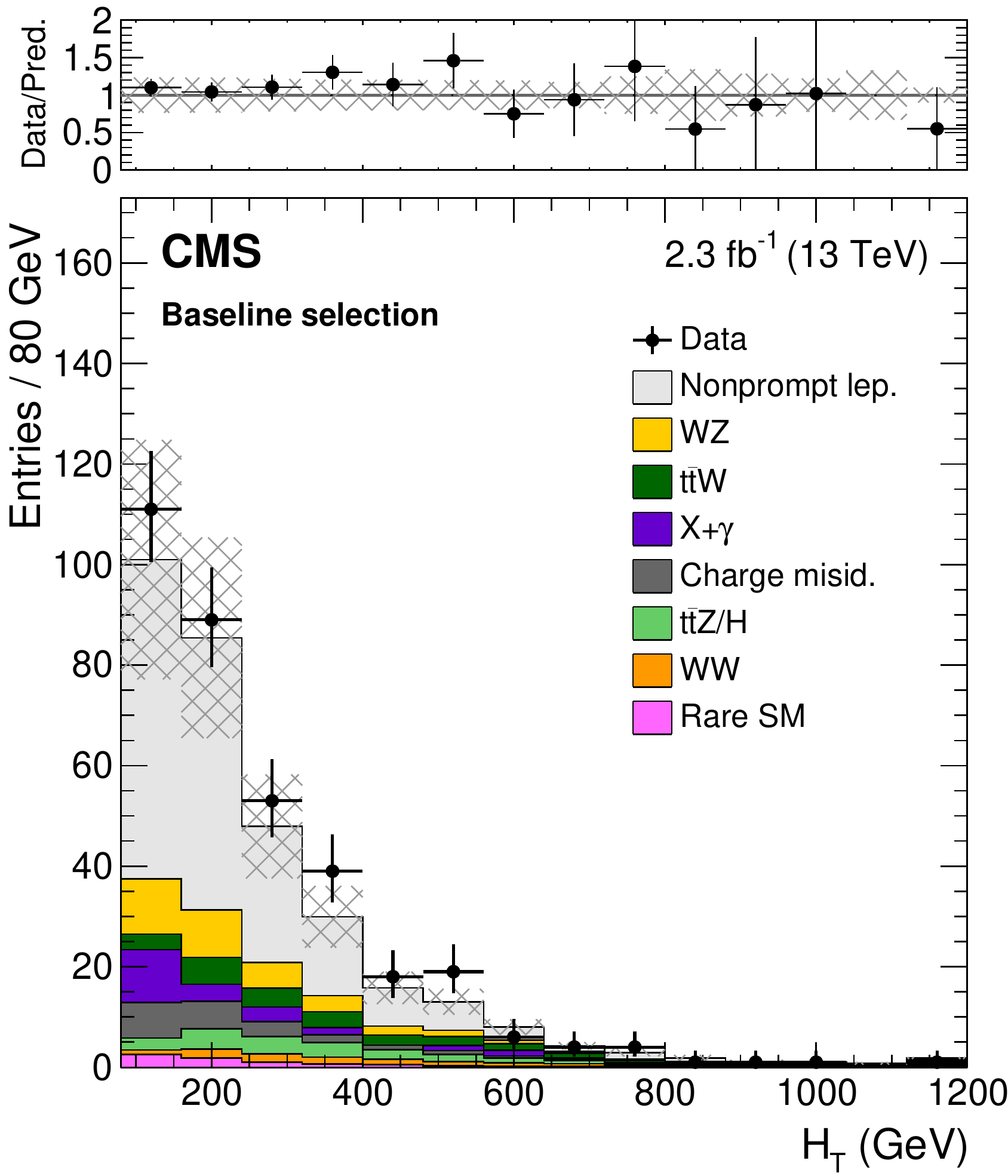}
\\
\includegraphics[width=.35\textwidth]{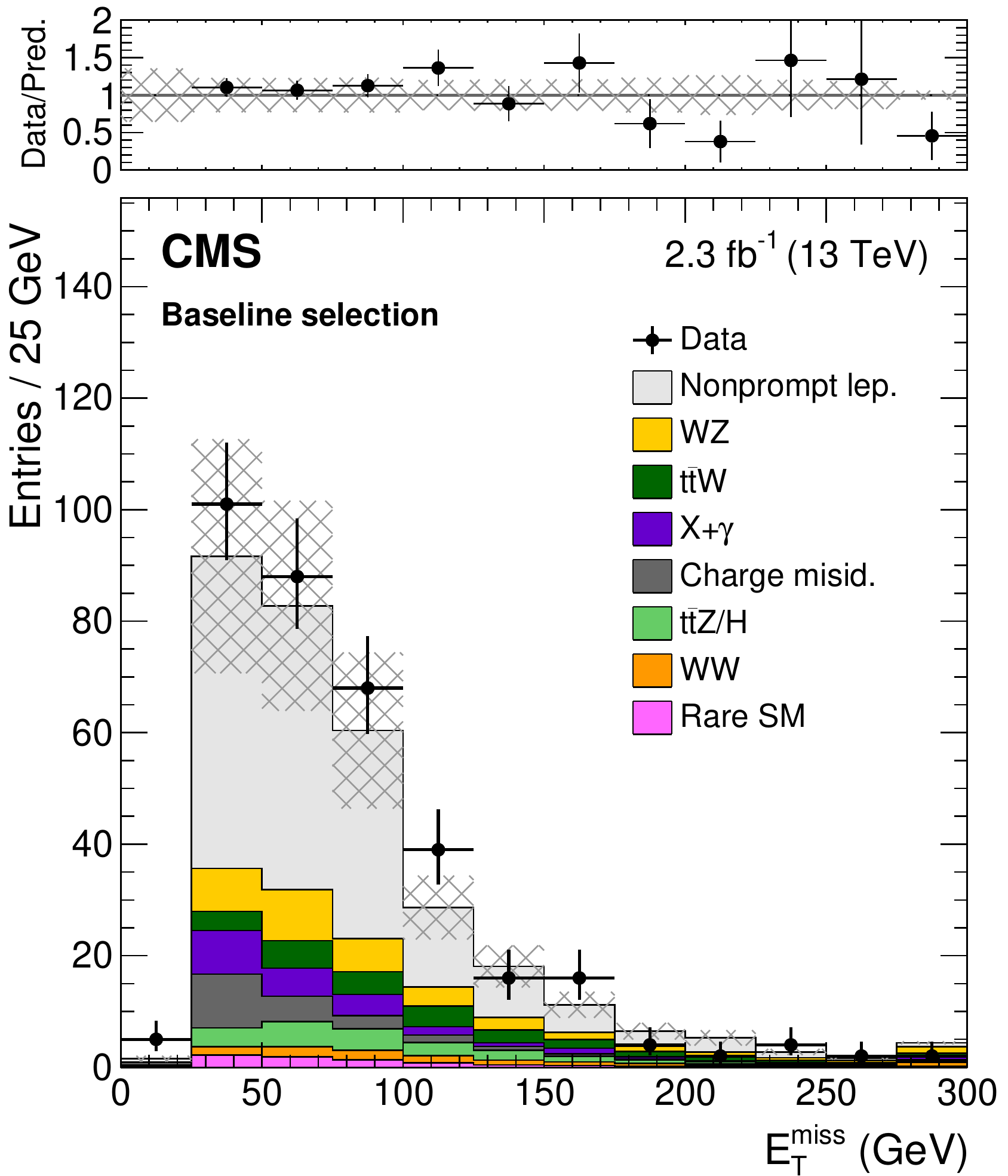}
\includegraphics[width=.35\textwidth]{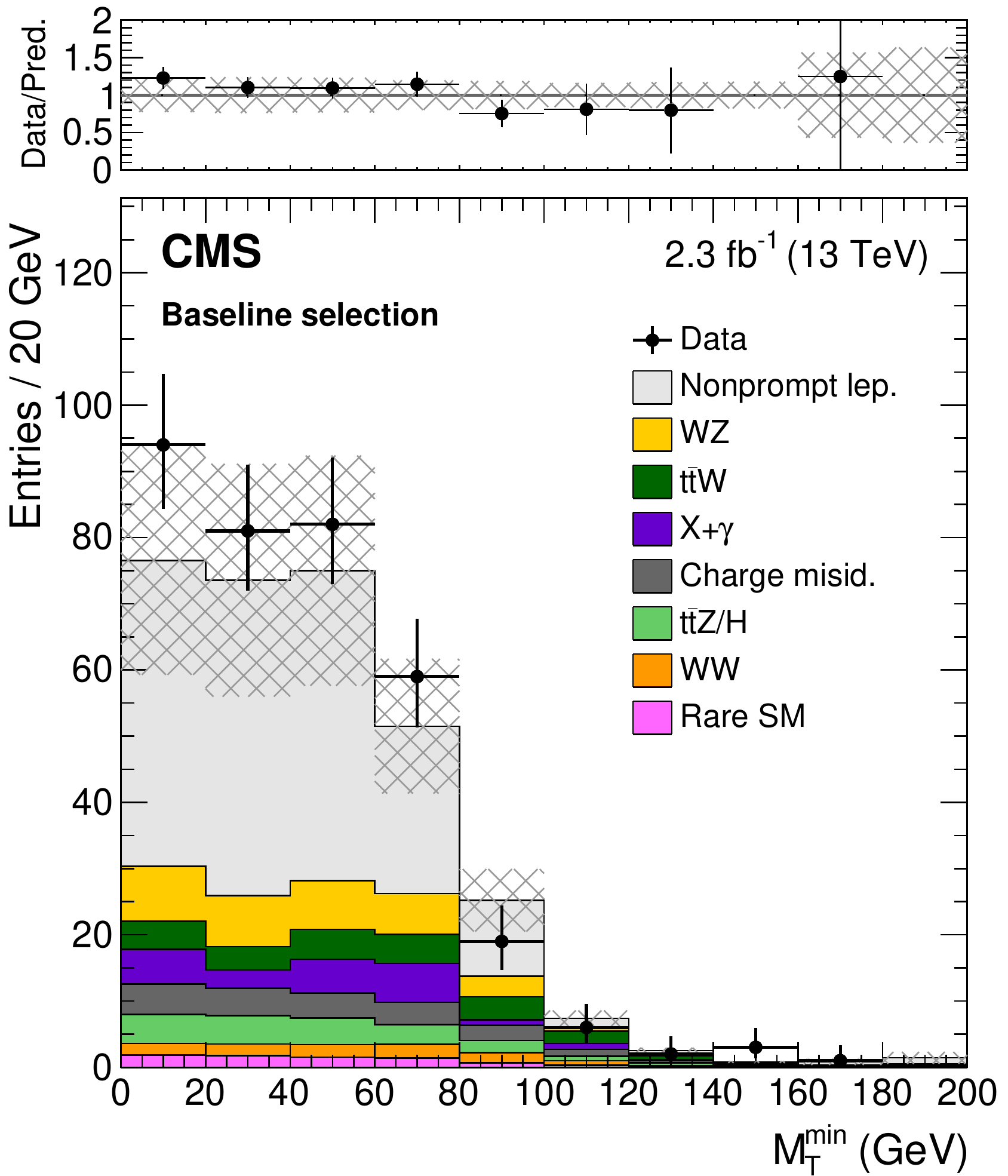}
\\
\includegraphics[width=.35\textwidth]{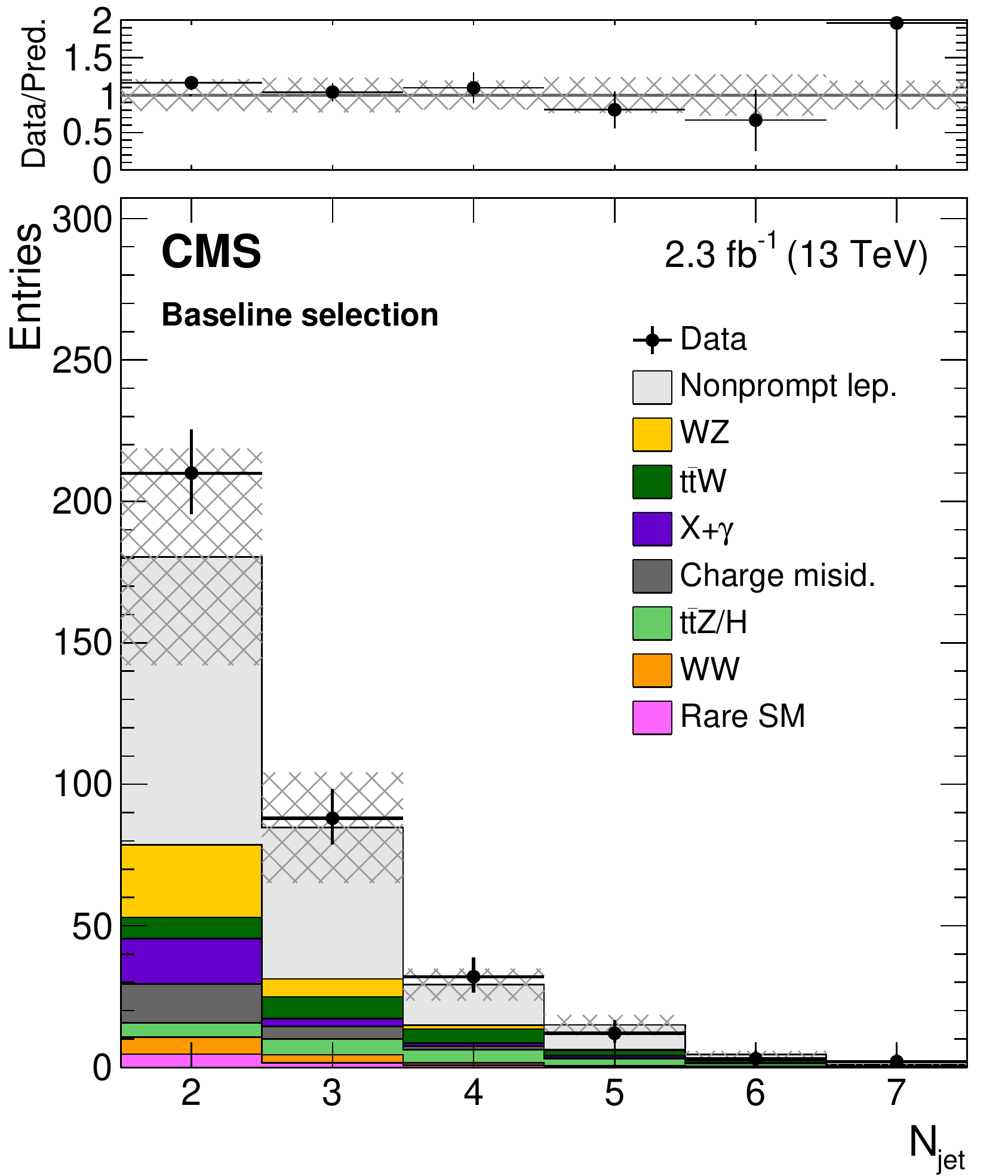}
\includegraphics[width=.35\textwidth]{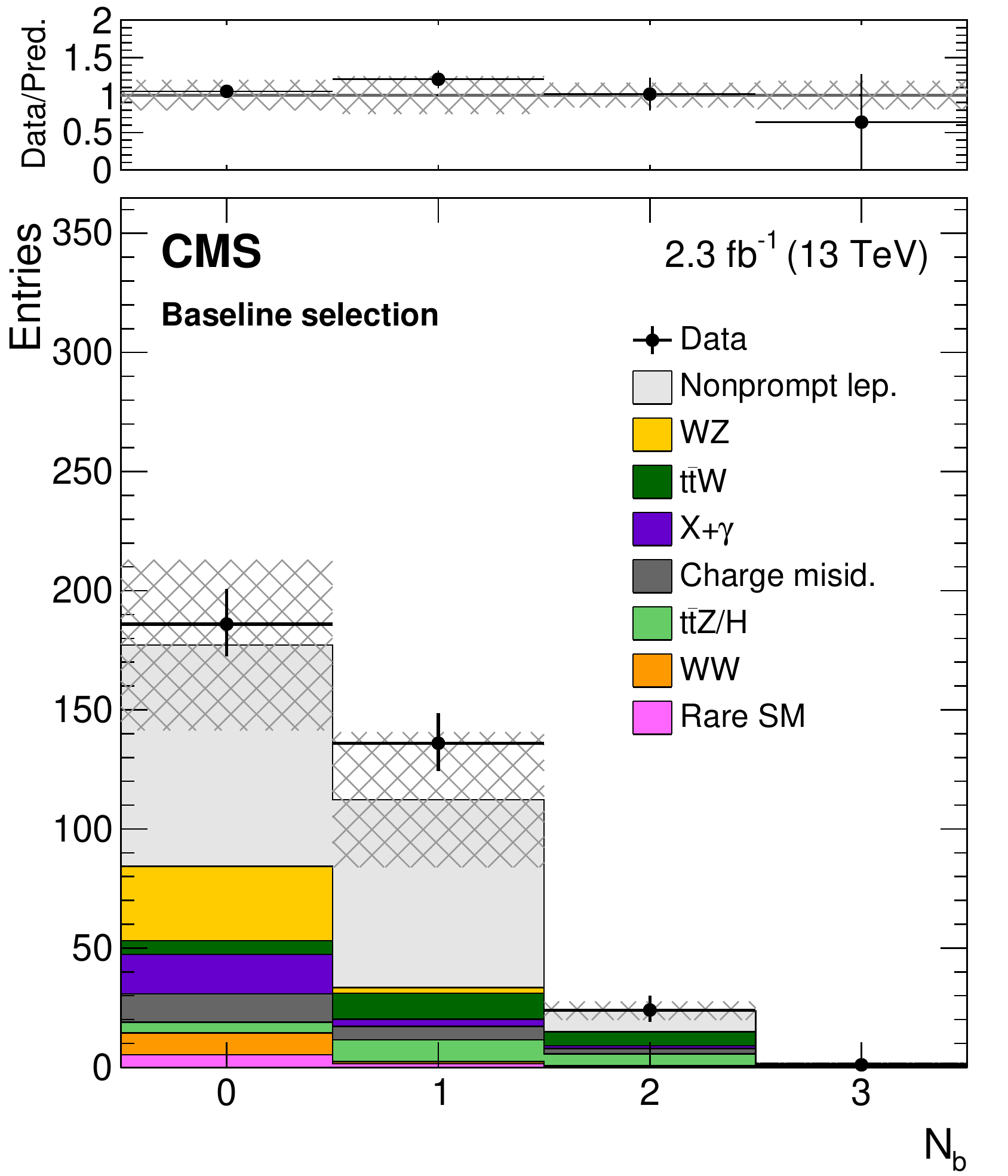}
\caption{Distributions of the main analysis variables: \HT~(top), \ETmiss~(middle left), \MTmin~(middle right), \Njets~(bottom left), and \Nbjets~(bottom right), after a baseline selection requiring a pair of SS leptons, two jets, and either $\ETmiss > 30\GeV$ or $\HT > 500\GeV$. The last bin includes the overflow. The notation X+$\gamma$ refers to processes with a prompt photon in the final state. The hatched area represents the total uncertainty in the background prediction. The upper panels show the ratio of the observed event yield to the background prediction.
}
\label{fig:kinem}
\end{figure*}

\begin{figure*}[!hbtp]
\centering
\includegraphics[width=.45\textwidth]{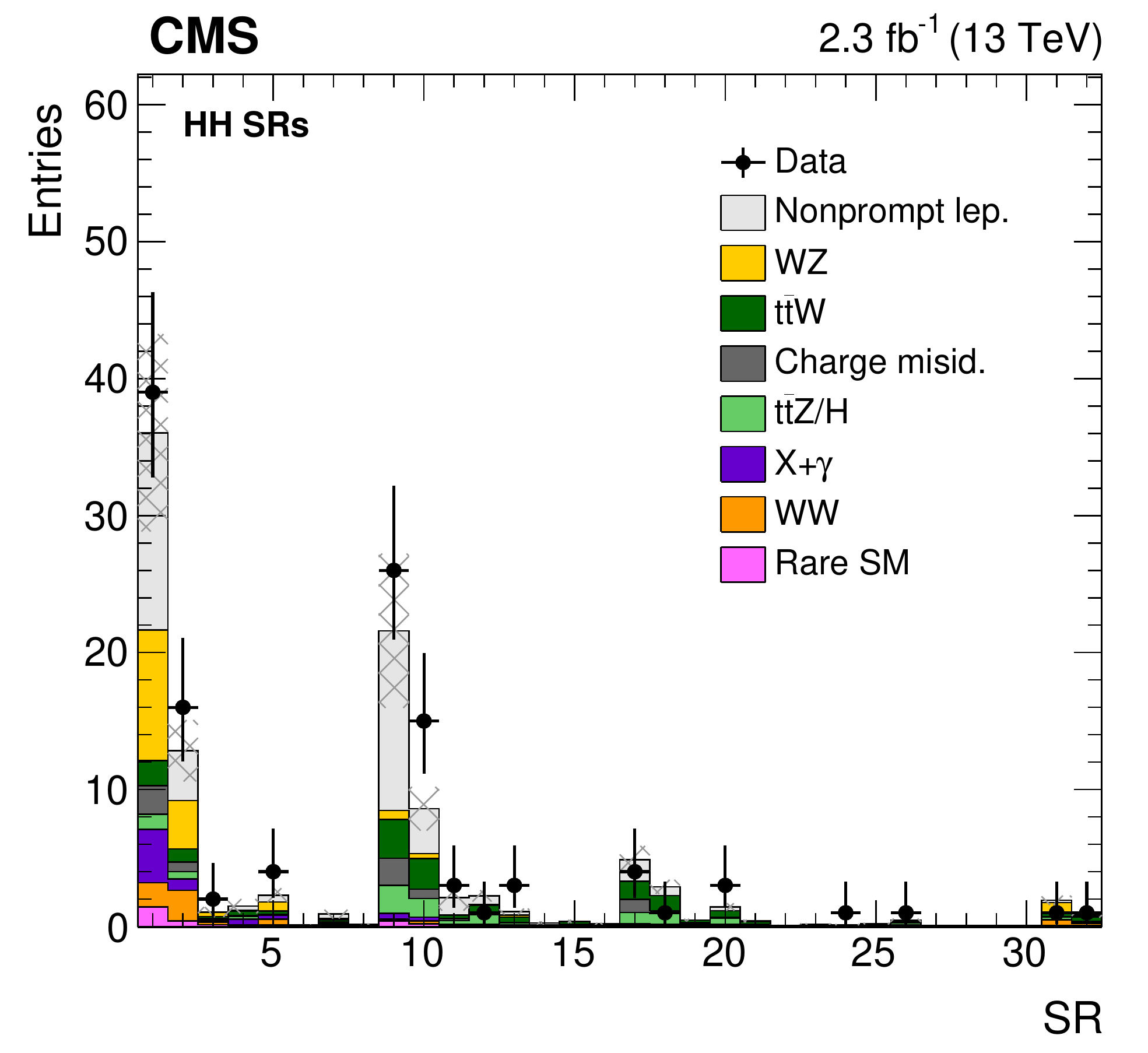}
\includegraphics[width=.45\textwidth]{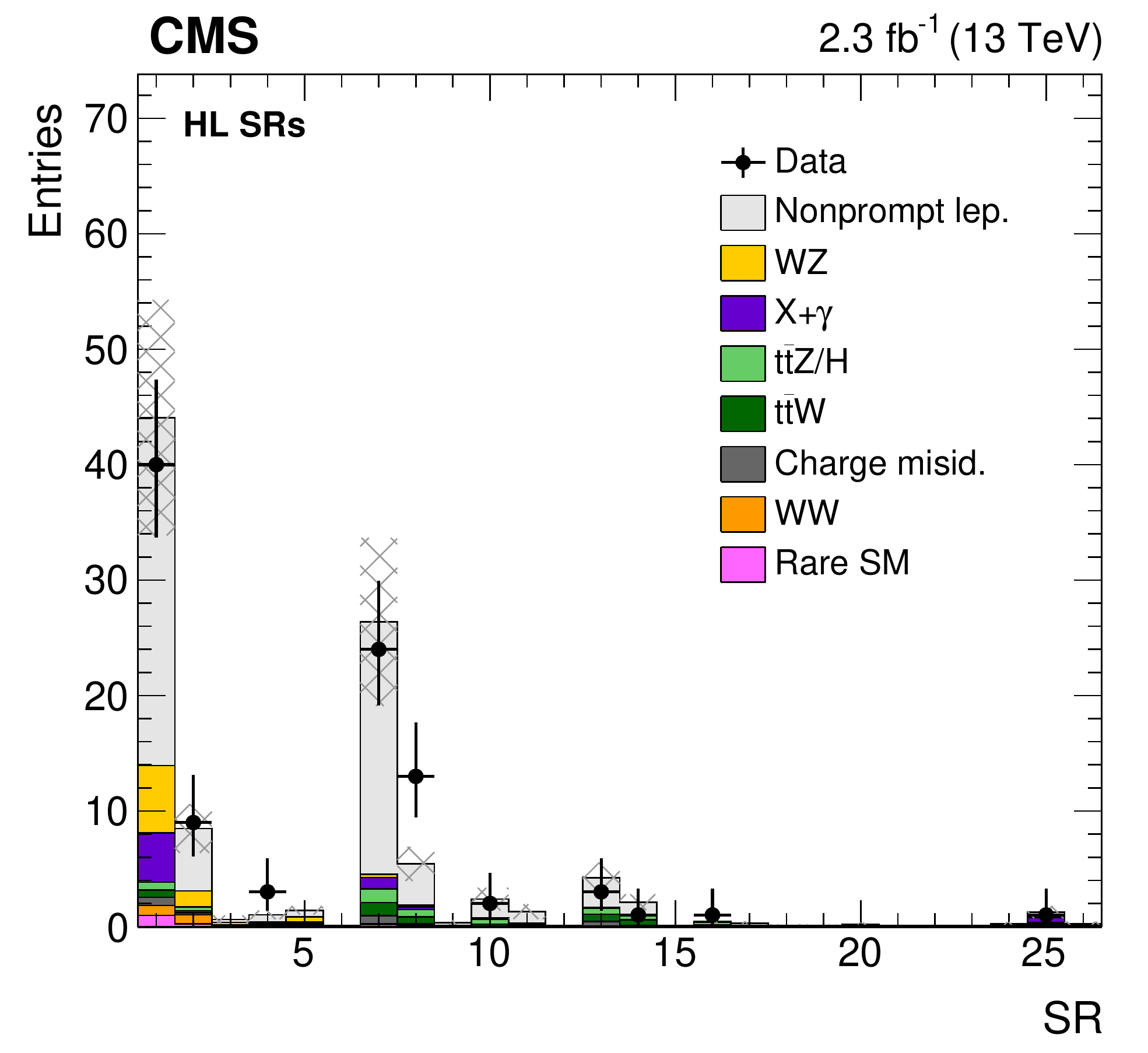}
\\
\includegraphics[width=.45\textwidth]{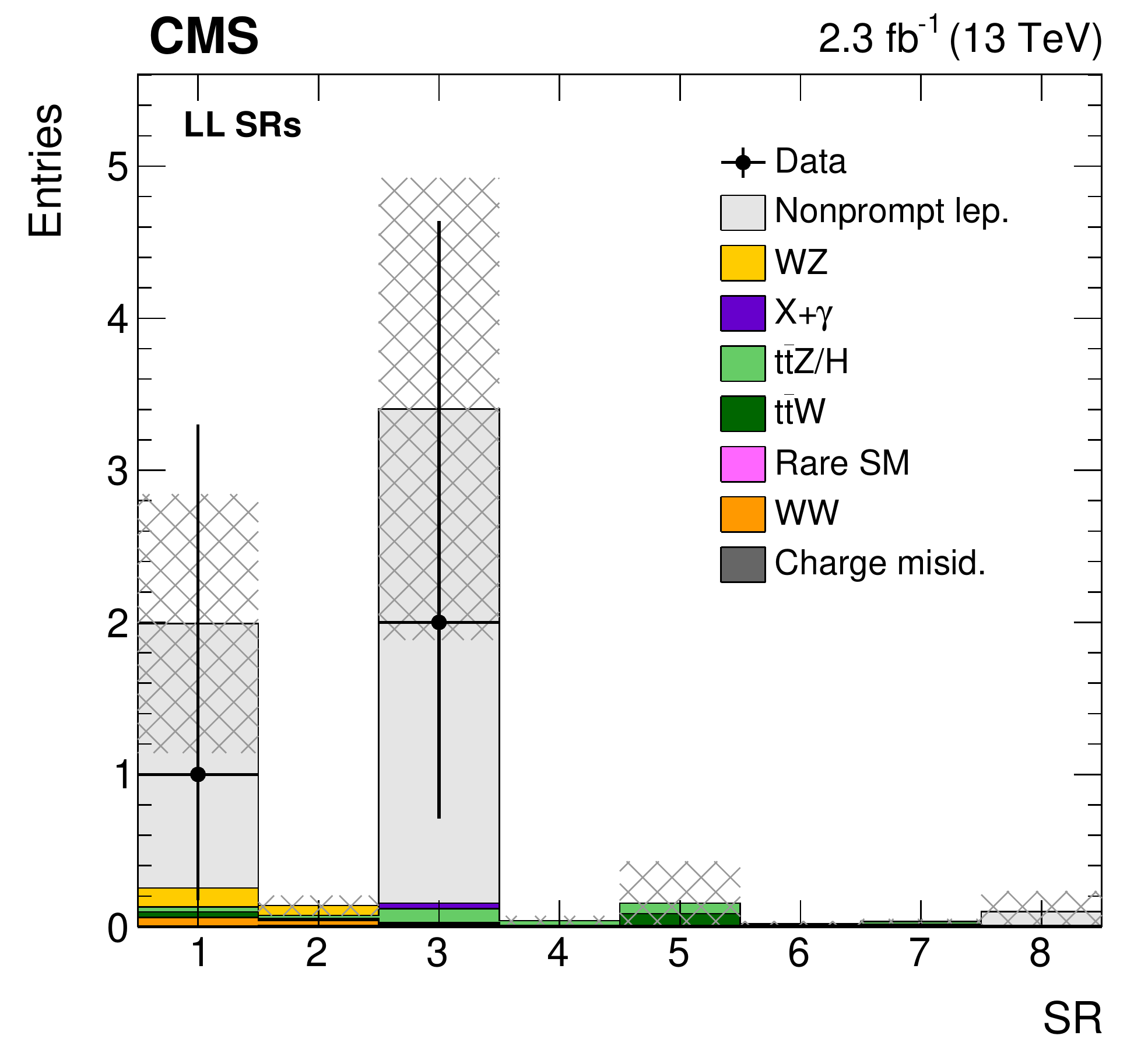}
\caption{Event yields in the HH~(top left), HL~(top right), and LL~(bottom) SRs. The notation X+$\gamma$ refers to processes with a prompt photon in the final state. The hatched area represents the total uncertainty in the background prediction.}
\label{fig:SR}
\end{figure*}

Distributions of the five kinematic variables used to define the SRs, \HT, \ETmiss, \MTmin, \Njets, and \Nbjets,
are shown in Fig.~\ref{fig:kinem} after a baseline selection requiring a pair of SS leptons, two jets, and either $\ETmiss > 30\GeV$ or $\HT > 500\GeV$.
The results are shown in comparison to the background prediction.
The event yields in the SRs after the full selection are presented in Fig.~\ref{fig:SR} and in Table~\ref{tab:yieldsSR};
no significant deviation from the SM background prediction is observed.
The largest local significances are 2.2 and 1.8 standard deviations in HL SR8 and in HH SR10, respectively.

\begin{table*}[!hbtp]
\centering
\topcaption{Expected number of background and observed events for the different SRs considered in this analysis.}
\label{tab:yieldsSR}
\cmsTable{
\begin{tabular}{c|cc|cc|cc}
\hline
     & \multicolumn{2}{c|}{HH event yields} & \multicolumn{2}{c|}{HL event yields} & \multicolumn{2}{c}{LL event yields} \\
Region & Expected SM & Observed  & Expected SM & Observed  & Expected SM & Observed \\
\hline
SR1  & 36.0 $\pm$ 7.0\x & 39  & 44.1 $\pm$ 10.9 & 40 &  1.99 $\pm$  0.94 &  1 \\
SR2  & 12.8 $\pm$ 2.1\x & 16  & 8.5 $\pm$ 2.1 & 9 &  0.14 $\pm$  0.07 &  0 \\
SR3  & 1.05 $\pm$ 0.36 & 2  & 0.61 $\pm$ 0.36 & 0 &  3.4 $\pm$  1.5 &  2 \\
SR4  & 1.49 $\pm$ 0.52 & 0  & 1.01 $\pm$ 0.38 & 3 &  0.04 $\pm$  0.03 &  0 \\
SR5  & 2.29 $\pm$ 0.49 & 4  & 1.40 $\pm$ 0.37 & 0 &  0.15 $\pm$  0.28 &  0 \\
SR6  & 0.11 $\pm$ 0.04 & 0  & 0.08 $\pm$ 0.04 & 0 &  0.02 $\pm$  0.01 &  0 \\
SR7  & 0.91 $\pm$ 0.31 & 0  & 26.4 $\pm$ 7.6\x & 24 &  0.03 $\pm$  0.01 &  0 \\
SR8  & 0.16 $\pm$ 0.06 & 0  & 5.4 $\pm$ 1.5 & 13 &  0.10 $\pm$  0.10 &  0 \\
\hline
SR9  & 21.6 $\pm$ 5.2\x & 26  & 0.34 $\pm$ 0.20 & 0 &  &   \\
SR10  & 8.6 $\pm$ 1.4 & 15  & 2.37 $\pm$ 0.99 & 2 &  &   \\
SR11  & 2.10 $\pm$ 0.92 & 3  & 1.29 $\pm$ 0.65 & 0 &  &   \\
SR12  & 2.24 $\pm$ 0.40 & 1  & 0.05 $\pm$ 0.04 & 0 &  &   \\
SR13  & 1.09 $\pm$ 0.21 & 3  & 4.2 $\pm$ 1.3 & 3 &  &   \\
SR14  & 0.25 $\pm$ 0.11 & 0  & 2.11 $\pm$ 0.69 & 1 &  &   \\
SR15  & 0.37 $\pm$ 0.12 & 0  & 0.06 $\pm$ 0.03 & 0 &  &   \\
SR16  & 0.19 $\pm$ 0.08 & 0  & 0.42 $\pm$ 0.09 & 1 &  &   \\
SR17  & 4.9 $\pm$ 1.0 & 4  & 0.29 $\pm$ 0.15 & 0 &  &   \\
SR18  & 2.90 $\pm$ 0.47 & 1  & 0.10 $\pm$ 0.08 & 0 &  &   \\
SR19  & 0.47 $\pm$ 0.09 & 0  & 0.11 $\pm$ 0.06 & 0 &  &   \\
SR20  & 1.43 $\pm$ 0.25 & 3  & 0.18 $\pm$ 0.17 & 0 &  &   \\
SR21  & 0.40 $\pm$ 0.10 & 0  & 0.001 $\pm$ 0.001 & 0 &  &   \\
SR22  & 0.08 $\pm$ 0.04 & 0  & 0.04 $\pm$ 0.04 & 0 &  &   \\
SR23  & 0.17 $\pm$ 0.06 & 0  & 0.03 $\pm$ 0.03 & 0 &  &   \\
SR24  & 0.14 $\pm$ 0.04 & 1  & 0.21 $\pm$ 0.17 & 0 &  &   \\
SR25  & 0.21 $\pm$ 0.06 & 0  & 1.25 $\pm$ 0.53 & 1 &  &   \\
SR26  & 0.46 $\pm$ 0.12 & 1  & 0.25 $\pm$ 0.12 & 0 &  &   \\
\hline
SR27  & 0.005 $\pm$ 0.016 & 0  & & &  &  \\
SR28  & 0.03 $\pm$ 0.02 & 0  & & &  &  \\
SR29  & 0.02 $\pm$ 0.01 & 0  & & &  &  \\
SR30  & 0.02 $\pm$ 0.01 & 0  & & &  &  \\
SR31  & 1.91 $\pm$ 0.32 & 1  & & &  &  \\
SR32  & 0.85 $\pm$ 0.18 & 1  & & &  &  \\
\hline
\end{tabular}}
\end{table*}

The results of the search are used to constrain
the benchmark SUSY models presented in Section~\ref{sec:strategy}.
For each mass point in the SUSY particle mass spectrum, results
from all SRs are combined to extract cross section exclusion
limits at the 95\% confidence level (CL), using the asymptotic formulation of
the modified frequentist CL$_\mathrm{s}$ criterion~\cite{Read:2002hq,Junk:1999kv,ATL-PHYS-PUB-2011-011,Cowan:2010js}.
Signal and background uncertainties are included as log-normal nuisance parameters and, when relevant, take into account correlation effects among different SRs and/or different processes.
Exclusion contours make use of the cross section values calculated at NLO plus next-to-leading logarithmic (NLL) accuracy,
assuming that all SUSY particles other than those included in the respective diagram are too heavy to participate in
the interaction~\cite{bib-nlo-nll-01,bib-nlo-nll-02,bib-nlo-nll-03,bib-nlo-nll-04,bib-nlo-nll-05,Borschensky:2014cia}.
In general, the SR with the largest sensitivity is HH SR31,
which requires $\ETmiss>300\GeV$ and is inclusive in the other variables.
However, depending on the model and the region of parameter space,
other SRs contribute significantly to the total sensitivity:
for instance, a considerable contribution comes from HL SR25 in case of signal models with a soft lepton,
from HH SR32 and HL SR26 in case of high \HT, from HH SR3 and SR8 in case of no b jets, and from HH SR24 and SR26 in case of 2 or more b jets.

\begin{figure*}[!hbtp]
\centering
\includegraphics[width=.45\textwidth]{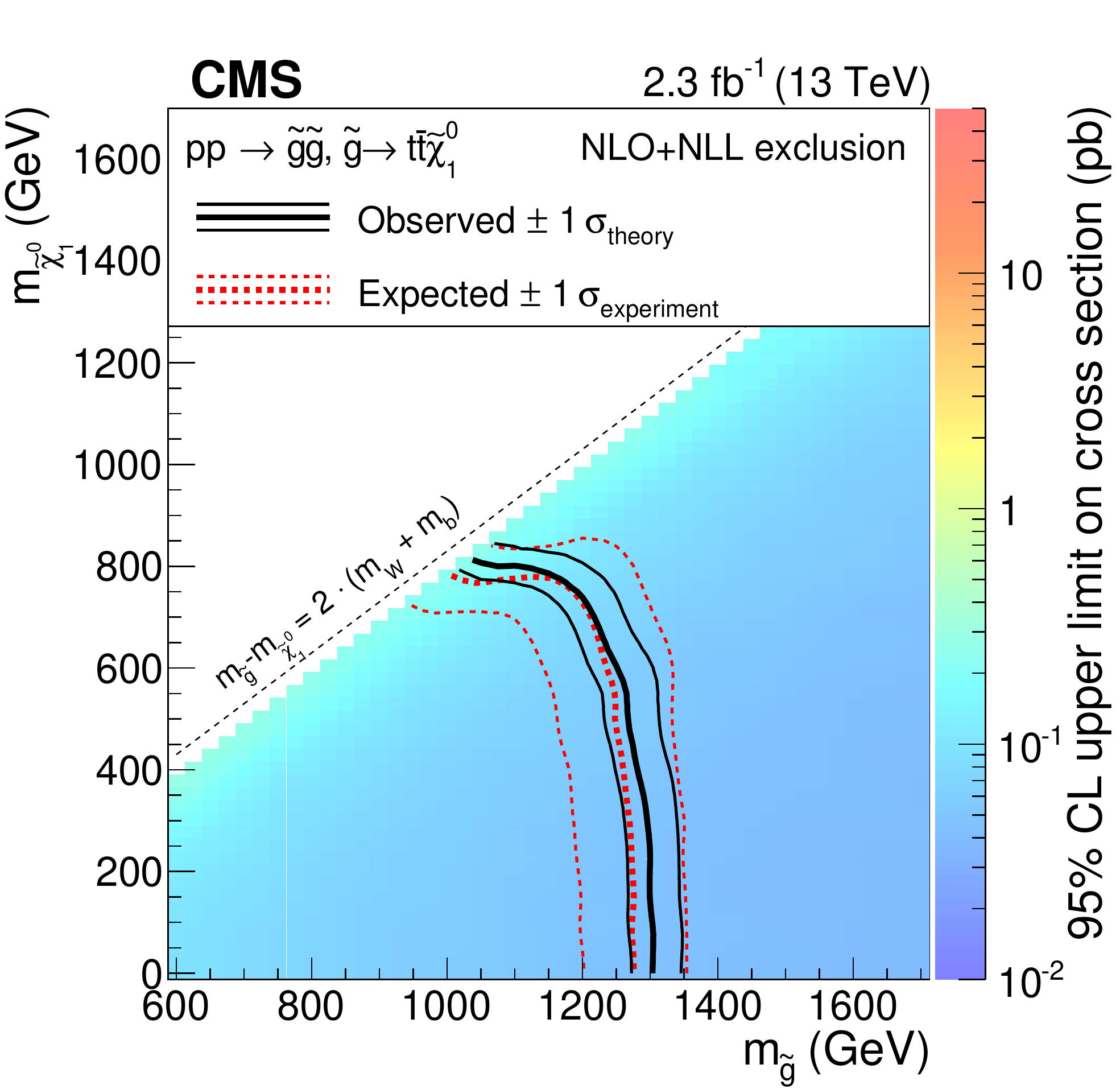}
\includegraphics[width=.45\textwidth]{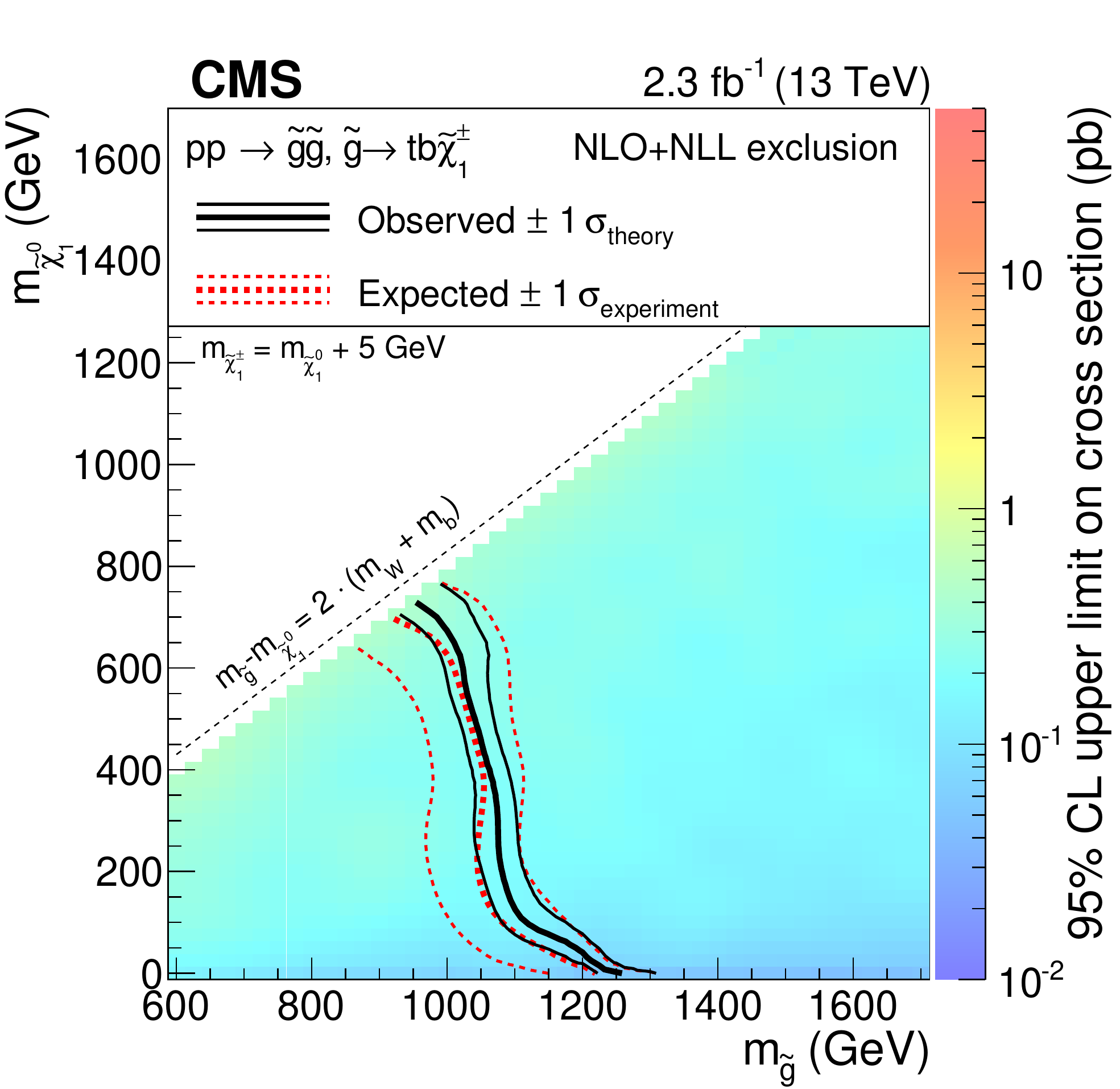}
\caption{Exclusion regions at the 95\% CL in the $m_{\PSGczDo}$ versus $m_{\PSg}$ plane for the \Totttt~(left) and \TfttbbWW~(right) models, where for the \TfttbbWW model $m_{\PSGcpmDo} = m_{\PSGczDo} + 5\GeV$. The right-hand side color scale indicates the excluded cross section values for a given point in the SUSY particle mass plane. The solid, black curves represent the observed exclusion limits assuming the NLO+NLL cross sections (thick line), or their variations of $\pm$1 standard deviation (thin lines). The dashed, red curves show the expected limits with the corresponding $\pm$1 standard deviation experimental uncertainties. Excluded regions are to the left and below the limit curves.}
\label{fig:t1ttxx_scan_xsec}
\end{figure*}

Results for models with gluinos decaying to virtual third generation squarks are shown in Fig.~\ref{fig:t1ttxx_scan_xsec} as a
function of the gluino and LSP masses.
For the \Totttt model (Fig.~\ref{fig:t1ttxx_scan_xsec}-left), in the regions of the SUSY parameter space with a large mass difference between the gluino and the LSP,
the results are rather stable with respect to LSP mass variations, and gluino masses up to 1300\GeV are excluded.
Near the kinematic threshold $m_{\PSg}-m_{\PSGczDo} = 2(m_{\PW}+m_{\PQb})$,
the gluino mass limit becomes weaker and is reduced to 1050\GeV for an LSP mass of 800\GeV.
Results for the \TfttbbWW model with nearly degenerate \PSGcpmDo and \PSGczDo masses are shown in Fig.~\ref{fig:t1ttxx_scan_xsec}-right;
the limit on the gluino mass lies in the range 950--1100\GeV except for very small \PSGcpmDo and \PSGczDo masses, where the sensitivity increases
because of the large Lorentz boost of the leptons from the \PSGcpmDo decay.

\begin{figure*}[!hbtp]
\centering
\includegraphics[width=.45\textwidth]{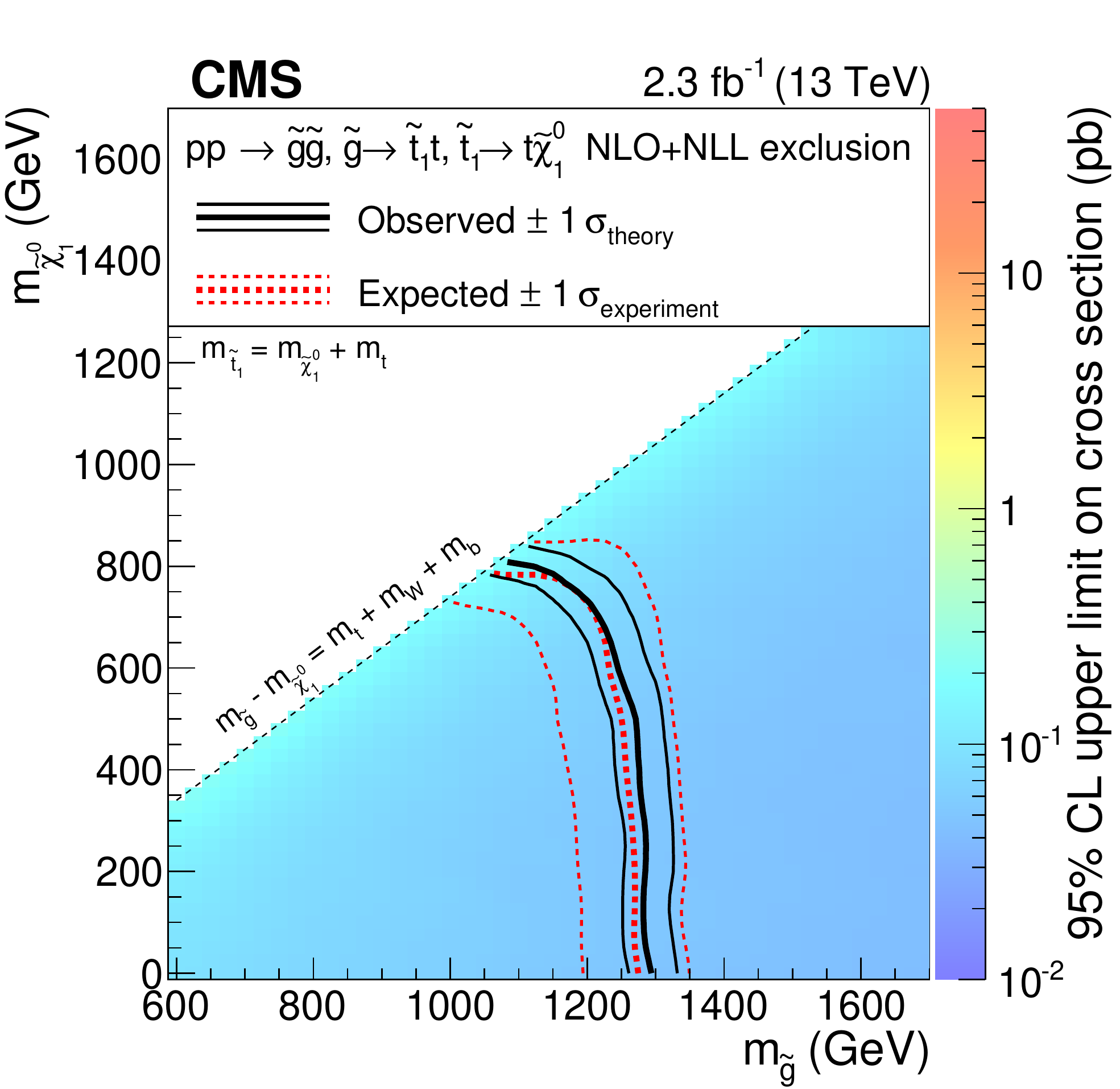}
\\
\includegraphics[width=.45\textwidth]{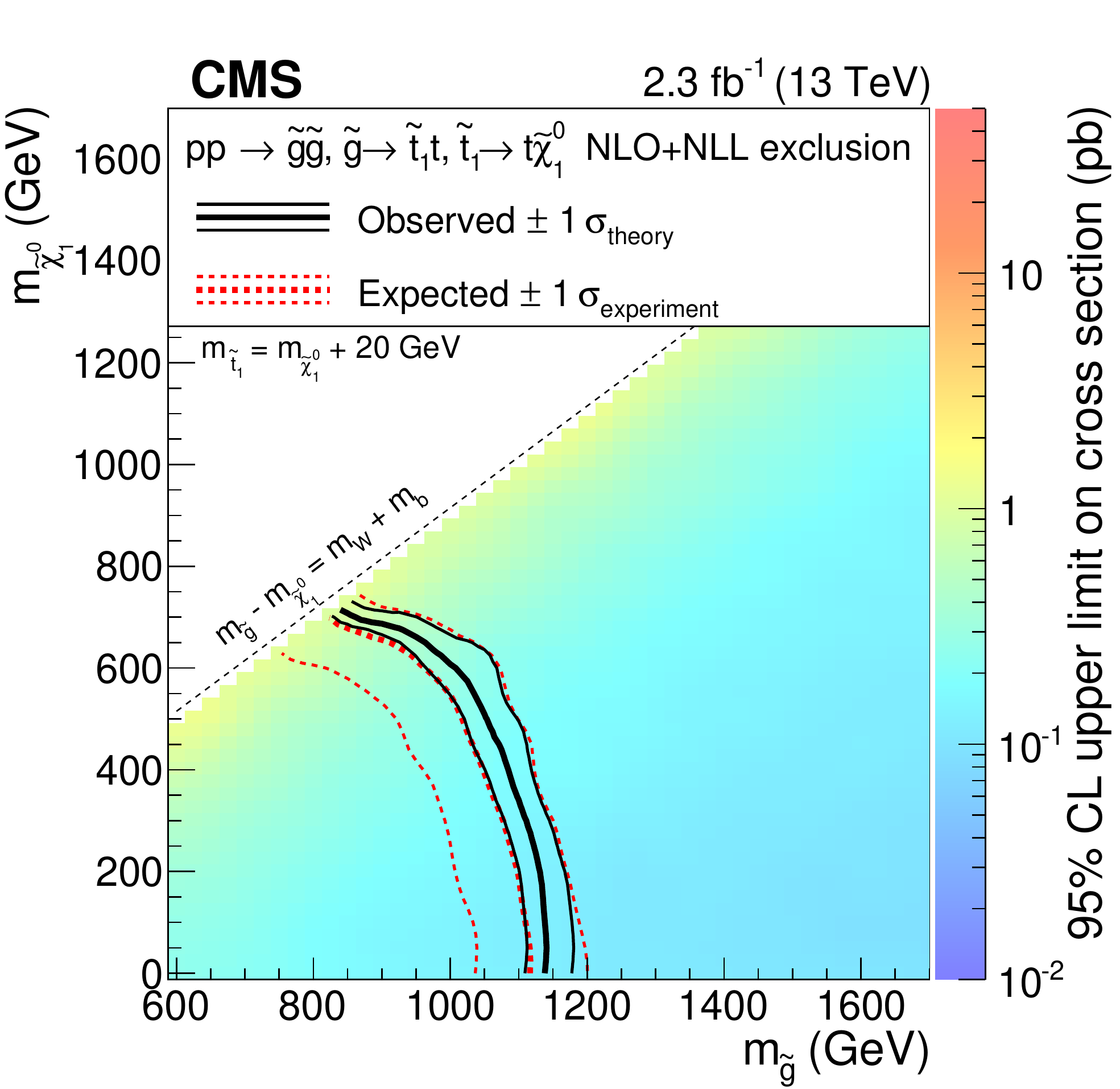}
\includegraphics[width=.45\textwidth]{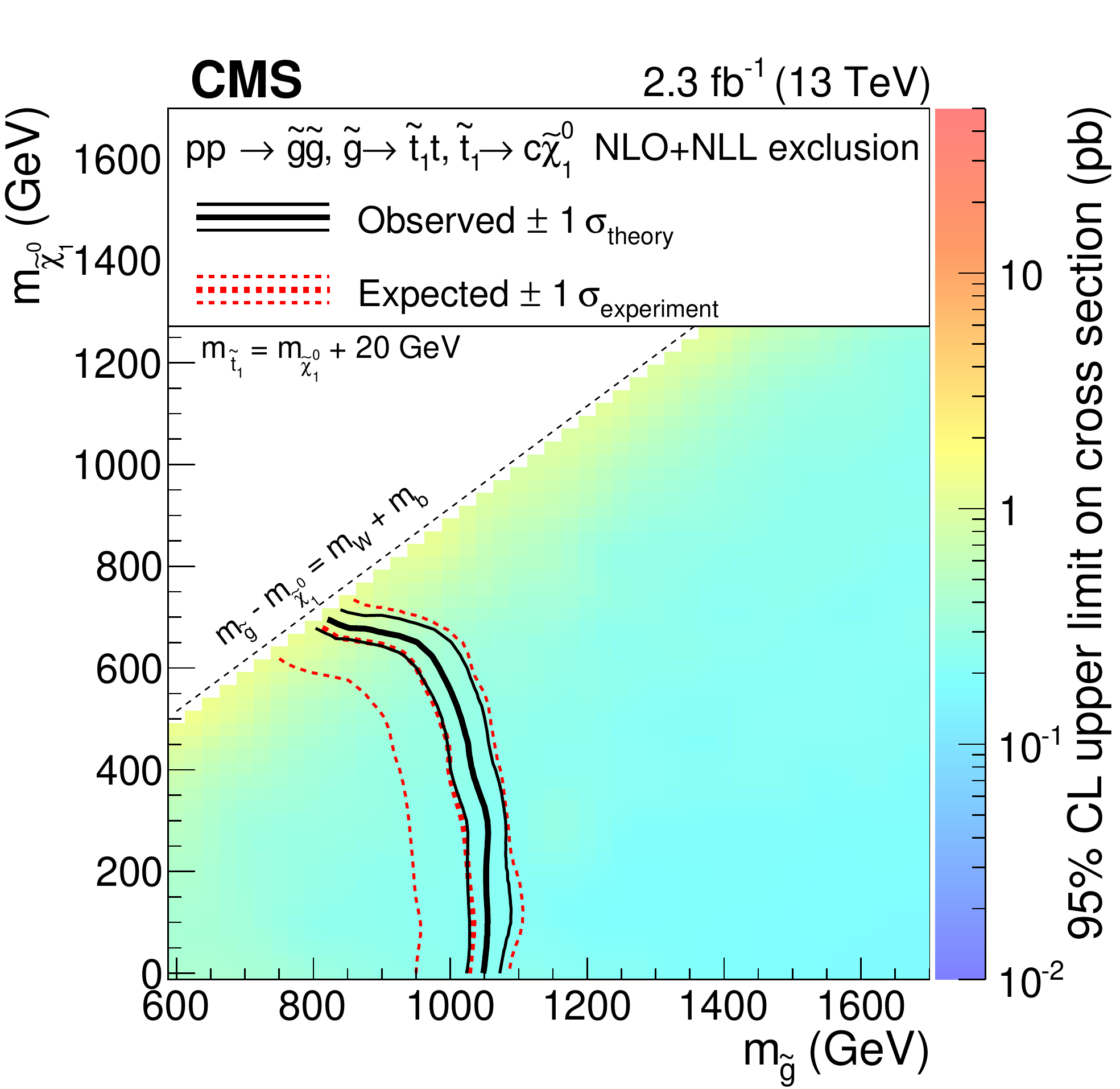}
\caption{Exclusion regions at the 95\% CL in the plane of $m_{\PSGczDo}$ versus $m_{\PSg}$ for models with the gluino decaying to an on-shell top squark: \Tftttt with $m_{\susytop} = m_{\PSGczDo} + m_{\PQt}$~(top), \Tftttt with $m_{\susytop} = m_{\PSGczDo} + 20\GeV$~(bottom left), and \Tfttcc with $m_{\susytop} = m_{\PSGczDo} + 20\GeV$~(bottom right). For a description of the notation, see Fig.~\ref{fig:t1ttxx_scan_xsec}.}
\label{fig:t5ttxx_scan_xsec}
\end{figure*}

Results for models with a gluino decaying to an on-shell top squark are shown in Fig.~\ref{fig:t5ttxx_scan_xsec} as a
function of the gluino and LSP masses.
For the \Tftttt model (Fig.~\ref{fig:t5ttxx_scan_xsec}-top), for which we take $m_{\susytop} = m_{\PSGczDo} + m_{\PQt}$,
similar exclusion curves are obtained as for the T1tttt model in Fig.~\ref{fig:t1ttxx_scan_xsec}-left because the production
cross section and the final-state particles are the same.
The limit becomes weaker when there is a small mass difference between the top squark and the LSP: for $m_{\susytop} - m_{\PSGczDo} = 20\GeV$,
the limit on the gluino mass is 1140\GeV for small LSP masses and about 850\GeV for $m_{\PSGczDo} = 700\GeV$ (Fig.~\ref{fig:t5ttxx_scan_xsec}-bottom left).
In the case of the \Tfttcc model with the same SUSY particle mass values, the sensitivity is slightly reduced because of the smaller number of leptons and b jets
in the final state (Fig.~\ref{fig:t5ttxx_scan_xsec}-bottom right).

\begin{figure*}[!hbtp]
\centering
\includegraphics[width=.45\textwidth]{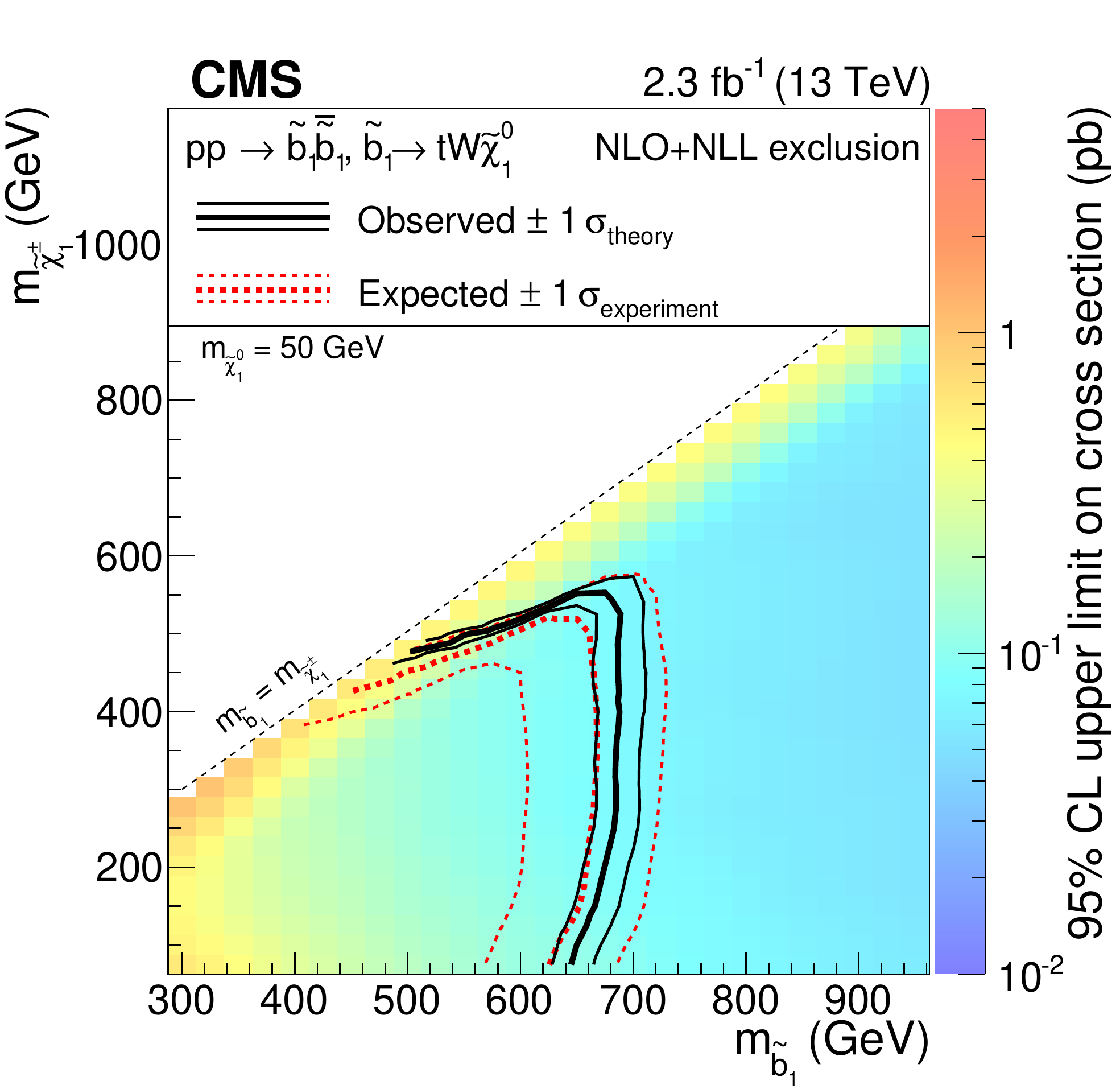}\\
\caption{Exclusion regions at the 95\% CL in the plane of $m_{\PSGcpmDo}$ versus $m_{\sbottom}$ for the \TsttWW model with $m_{\PSGczDo}=50\GeV$.
For a description of the notation, see Fig.~\ref{fig:t1ttxx_scan_xsec}.}
\label{fig:t6ttww_scan_xsec}
\end{figure*}

Figure~\ref{fig:t6ttww_scan_xsec} shows the results for b squark production in the \TsttWW model in the chargino (\PSGcpmDo) versus b squark mass plane,
where the LSP mass is assumed to be $m_{\PSGczDo}=50\GeV$. For chargino masses up to 550\GeV, b squark masses below 680\GeV are excluded.
The limit on the b squark mass is reduced to 500\GeV in regions where $m_{\PSGcpmDo}$ is within 100\GeV of $m_{\sbottom}$,
while a milder reduction is observed in regions where the difference between $m_{\PSGcpmDo}$ and $m_{\PSGczDo}$ is less than 150\GeV.

\begin{figure*}[!hbtp]
\centering
\includegraphics[width=.45\textwidth]{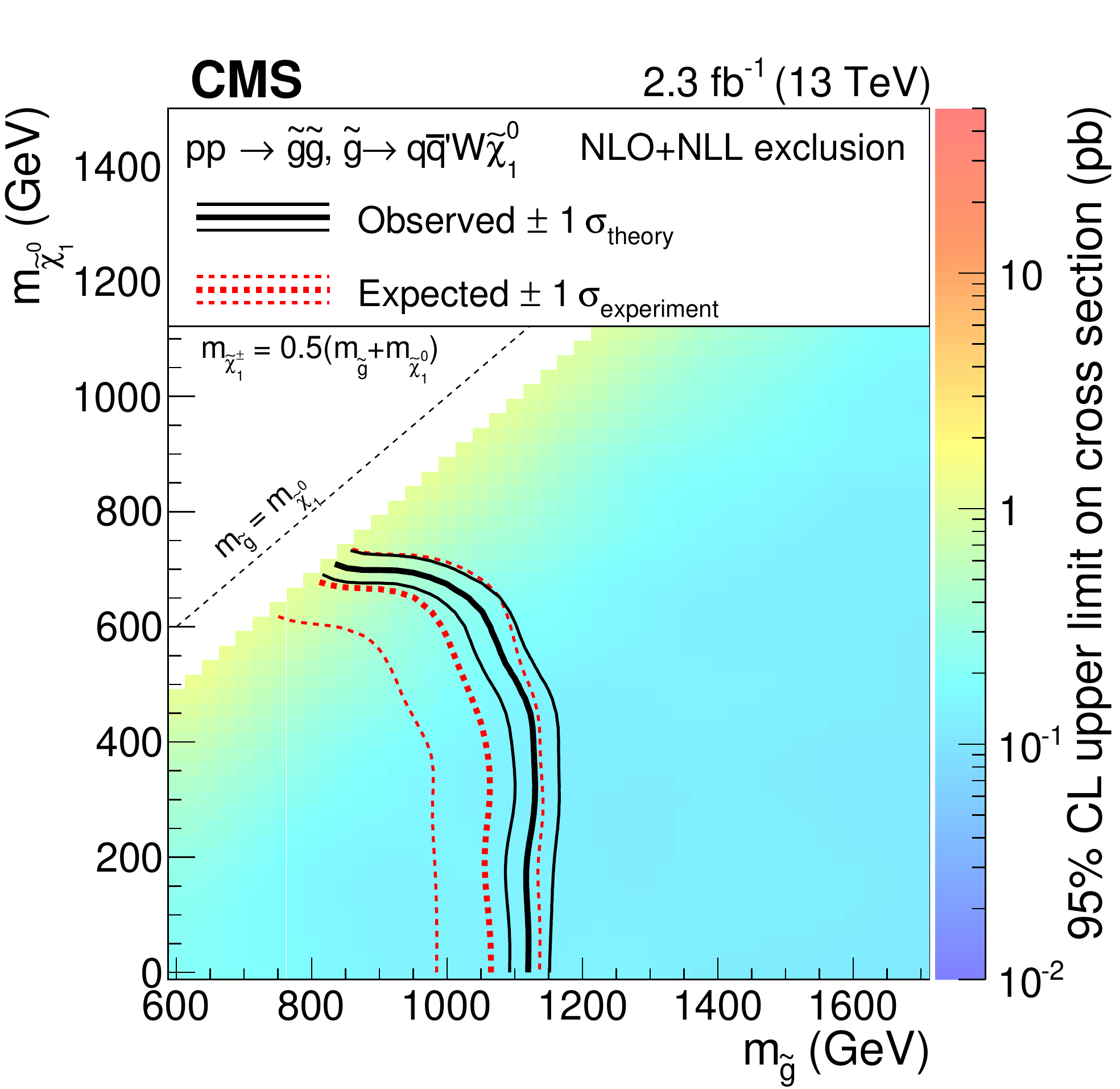}
\includegraphics[width=.45\textwidth]{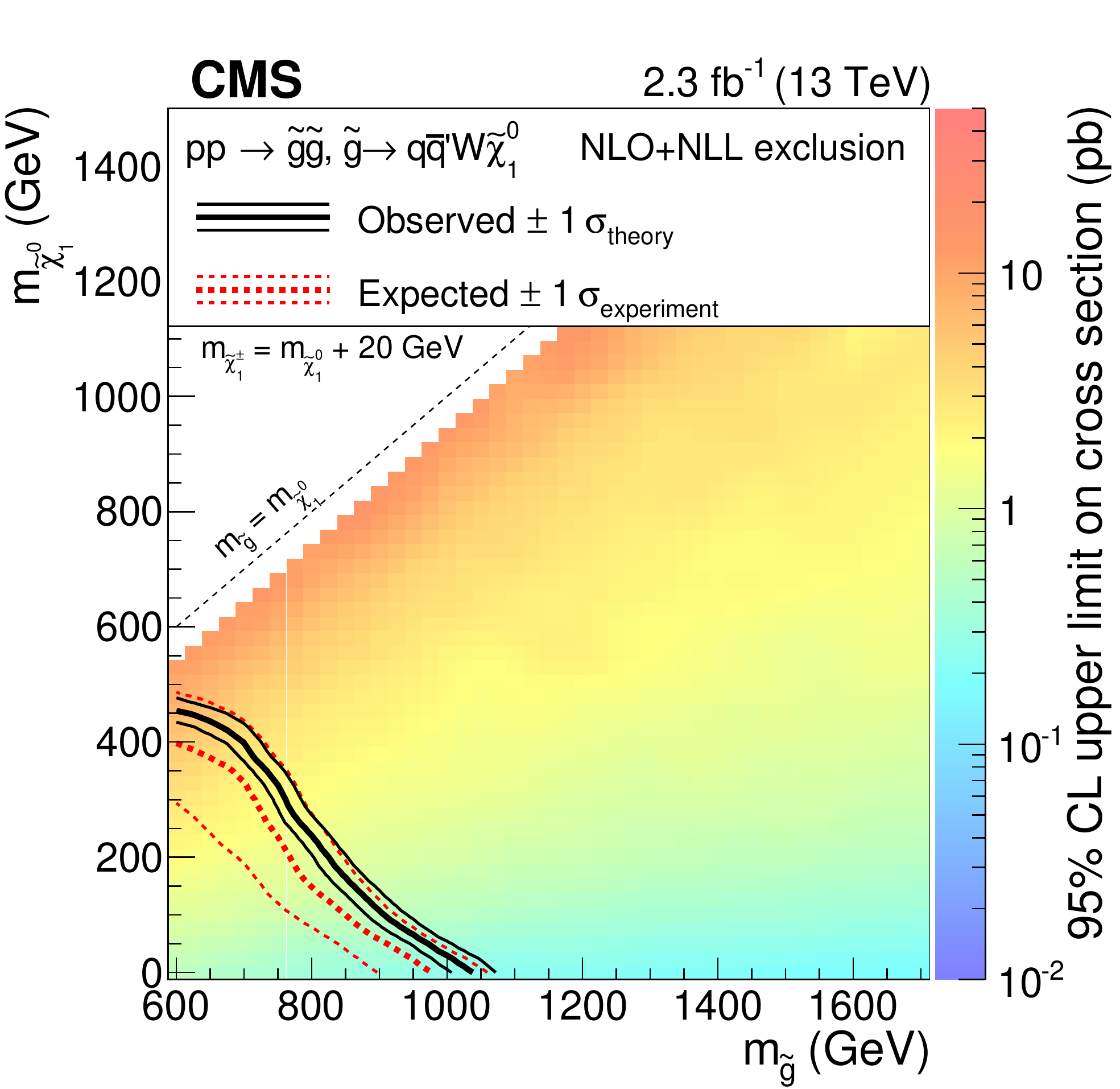}
\caption{Exclusion regions at the 95\% CL in the plane of $m_{\PSGczDo}$ versus $m_{\PSg}$ for the \TfqqqqWW model with $m_{\PSGcpmDo}=0.5(m_{\PSg} + m_{\PSGczDo})$~(left) and with $m_{\PSGcpmDo} = m_{\PSGczDo} + 20\GeV$~(right). For a description of the notation, see Fig.~\ref{fig:t1ttxx_scan_xsec}.}
\label{fig:t5qqqqww_scan_xsec}
\end{figure*}

Results for the \TfqqqqWW model are shown in Fig.~\ref{fig:t5qqqqww_scan_xsec} as a function of the gluino and LSP masses, with two different assumptions
for the chargino mass: it is either assumed to be the average of $m_{\PSg}$ and $m_{\PSGczDo}$, or it is set to $m_{\PSGczDo} + 20\GeV$.
In the first case (Fig.~\ref{fig:t5qqqqww_scan_xsec}-left), the exclusion limit on gluino masses exceeds 1100\GeV for LSP masses up to 400\GeV;
for larger LSP masses the limit is reduced to 830\GeV at $m_{\PSGczDo} = 700\GeV$.
In the second case (Fig.~\ref{fig:t5qqqqww_scan_xsec}-right), due to the smaller mass difference, leptons in the final state are soft and thus the sensitivity is reduced.

The results of the search are also used to set 95\% CL upper limits on the double \ttbar production cross section,
whose SM value computed at NLO precision~\cite{MADGRAPH5} is 9.1\fb.
The upper limit on $\sigma({\Pp\Pp} \to {\ttbar\ttbar})$ is found to be 119\fb, with an expected result of $102^{+57}_{-35}\fb$.
With the current integrated luminosity, the sensitivity to this signature is limited by the statistical precision.

Limits at the 95\% CL on the SS top quark pair production cross section are determined using events that satisfy the baseline selection categorized according to number of b jets (Fig.~\ref{fig:kinem}-bottom right);
apart from the charge requirement, the detector acceptance and the selection efficiency for the signal are assumed to match those of SM \ttbar events.
The observed (expected) upper limit on $\sigma(\Pp\Pp \to {\PQt\PQt}) + \sigma({\Pp\Pp} \to {\PAQt\PAQt})$ is 1.7\pb\,($1.5^{+0.7}_{-0.4}\pb$).

\begin{figure*}[!hbtp]
\centering
\includegraphics[width=.45\textwidth]{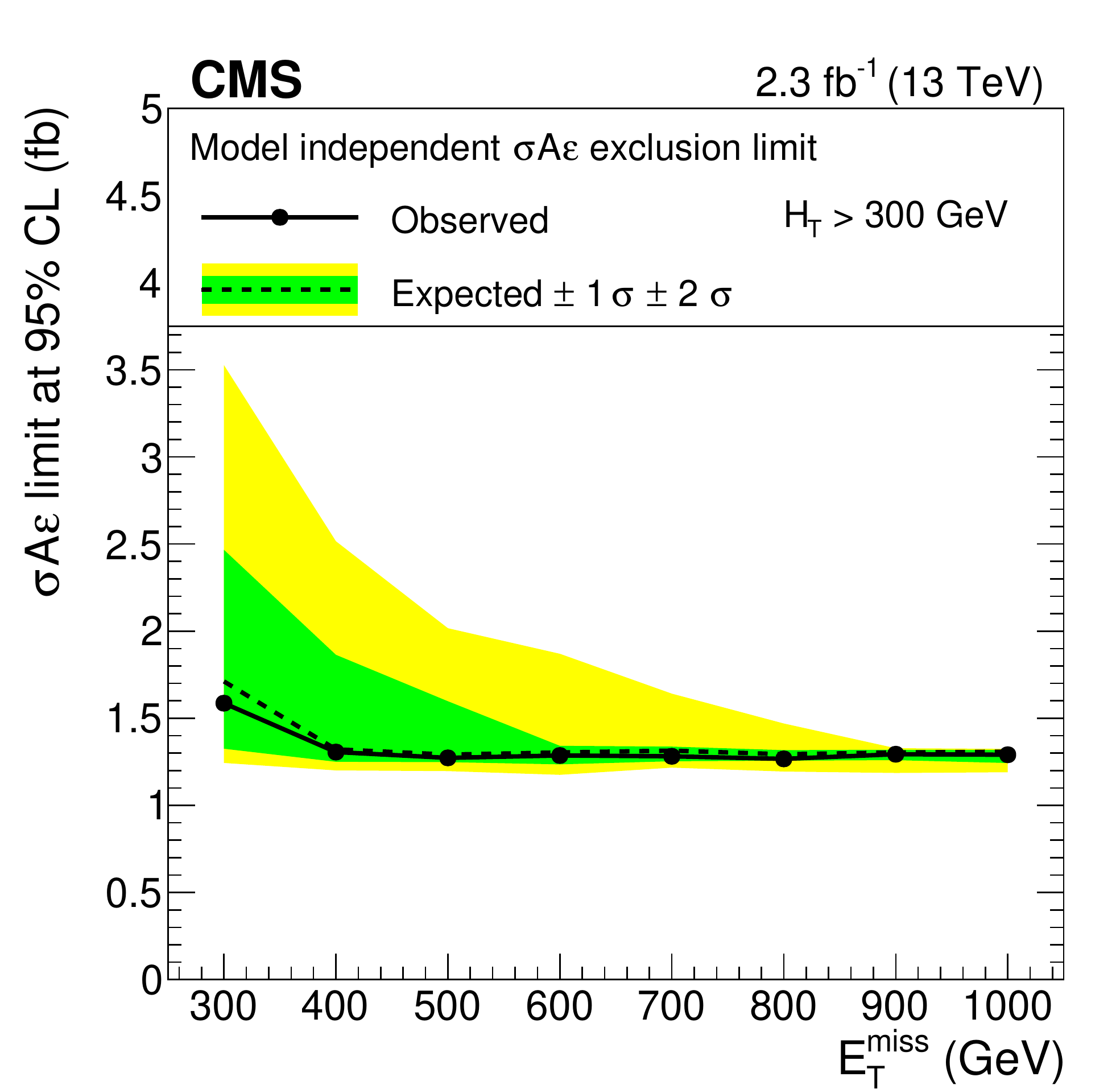}
\label{subfig:MIht}
\includegraphics[width=.45\textwidth]{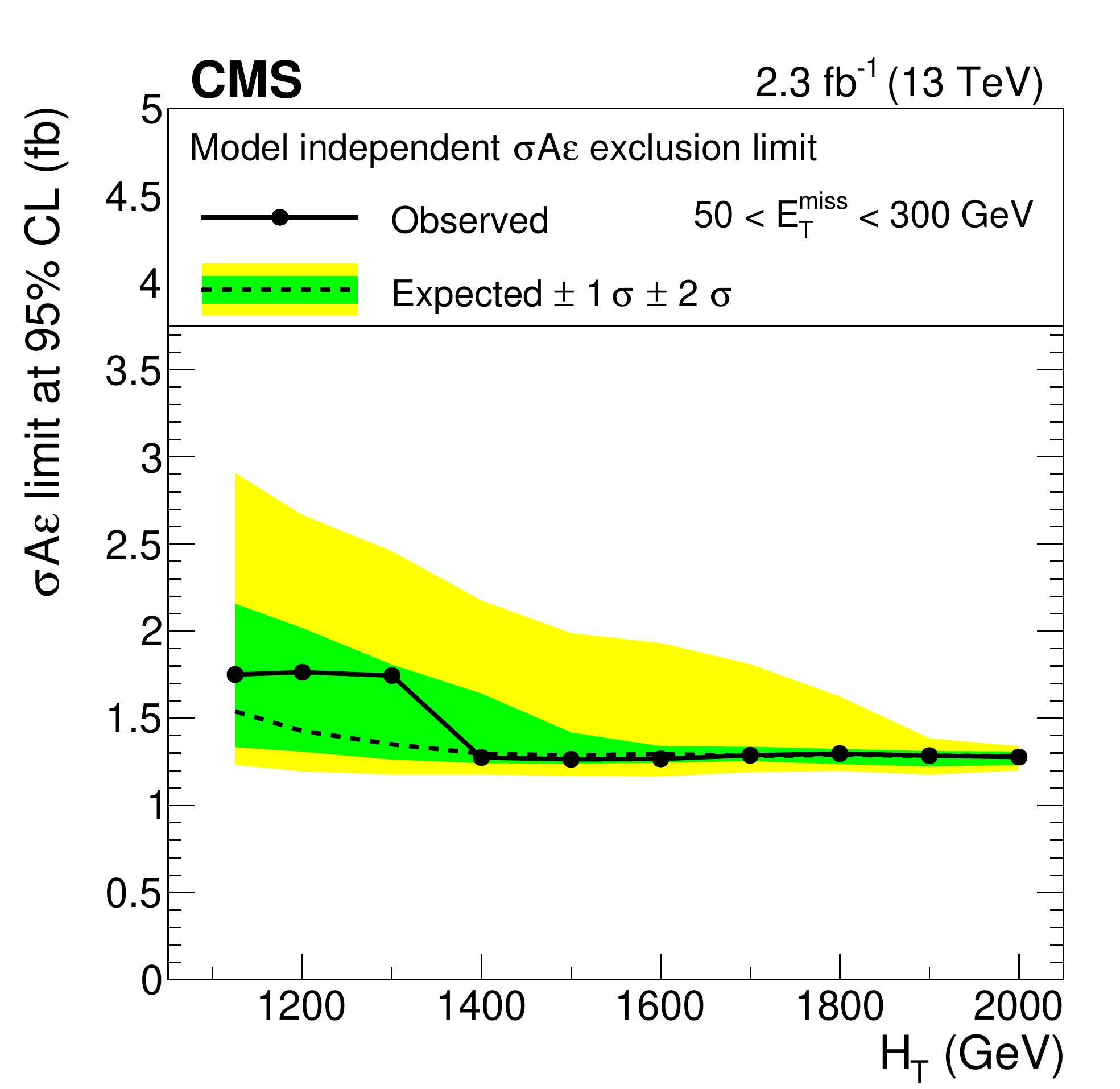}
\caption{Limits on the product of cross section, detector acceptance, and selection efficiency, $\sigma \! \mathcal{A} \epsilon$, for the production of an SS dilepton pair as a function of \MET in HH SR31~(left) and of \HT in HH SR32~(right).}
\label{fig:MIlimits}
\end{figure*}

Finally, we report model independent limits on the product of cross section, detector acceptance, and selection efficiency, $\sigma \! \mathcal{A} \epsilon$,
for the production of an SS dilepton pair in the two inclusive HH regions, SR31 and SR32,
using the CL$_\mathrm{s}$ criterion without the asymptotic approximation.
In SR31 the limit is computed as a function of the minimum threshold on \ETmiss for $\HT>300\GeV$, while in SR32
it is computed as a function of the \HT threshold for $50<\ETmiss<300\GeV$.
The results are shown in Fig.~\ref{fig:MIlimits}, where, in regions with no observed events, the minimum limit value of 1.3\fb is obtained.
These limits can be used to test additional BSM models, after accounting for the event selection efficiency.
The lepton efficiency ranges between 70--85\% (45--70\%) for generated muons (electrons)
with $\abs{\eta}<2.4$ and $\pt>25\GeV$, increasing as a function of \pt and converging to the maximum value for $\pt>60\GeV$;
the efficiencies of the \HT and \ETmiss requirements are mostly determined by the jet energy and \MET resolutions,
which are discussed in Refs.~\cite{Chatrchyan:2011ds,JME-13-003}.

\section{Summary}
\label{sec:summary}

The results of a search for new physics in same-sign dilepton events using the CMS detector at the LHC and
based on a data sample of pp collisions at $\sqrt{s} = 13\TeV$, corresponding to an integrated luminosity of \sslumi, are presented.
The data are analyzed in nonoverlapping signal regions defined with different selections
on lepton and event kinematic variables, as well as jet and b quark jet multiplicities.

No significant deviation from the standard model expectations is observed.
The results are used to set limits on the production of supersymmetric particles in various simplified models.
Gluino and bottom squark masses are excluded at the 95\% confidence level up to 1300 and 680\GeV, respectively.
These results extend the limits obtained in the previous version of the analysis~\cite{SUS-13-013}
by about 250\GeV on the gluino mass, and 150\GeV on the bottom squark mass.
In addition, 95\% confidence level upper limits of 119\fb and 1.7\pb are set on the cross sections for
the production of two top quark-antiquark pairs and for the production of two SS top quarks, respectively.
Model independent limits and selection efficiencies are provided to allow further interpretations of the results,
using alternative models to those examined here.

\begin{acknowledgments}

We congratulate our colleagues in the CERN accelerator departments for the excellent performance of the LHC and thank the technical and administrative staffs at CERN and at other CMS institutes for their contributions to the success of the CMS effort. In addition, we gratefully acknowledge the computing centres and personnel of the Worldwide LHC Computing Grid for delivering so effectively the computing infrastructure essential to our analyses. Finally, we acknowledge the enduring support for the construction and operation of the LHC and the CMS detector provided by the following funding agencies: BMWFW and FWF (Austria); FNRS and FWO (Belgium); CNPq, CAPES, FAPERJ, and FAPESP (Brazil); MES (Bulgaria); CERN; CAS, MoST, and NSFC (China); COLCIENCIAS (Colombia); MSES and CSF (Croatia); RPF (Cyprus); MoER, ERC IUT and ERDF (Estonia); Academy of Finland, MEC, and HIP (Finland); CEA and CNRS/IN2P3 (France); BMBF, DFG, and HGF (Germany); GSRT (Greece); OTKA and NIH (Hungary); DAE and DST (India); IPM (Iran); SFI (Ireland); INFN (Italy); MSIP and NRF (Republic of Korea); LAS (Lithuania); MOE and UM (Malaysia); BUAP, CINVESTAV, CONACYT, LNS, SEP, and UASLP-FAI (Mexico); MBIE (New Zealand); PAEC (Pakistan); MSHE and NSC (Poland); FCT (Portugal); JINR (Dubna); MON, RosAtom, RAS and RFBR (Russia); MESTD (Serbia); SEIDI and CPAN (Spain); Swiss Funding Agencies (Switzerland); MST (Taipei); ThEPCenter, IPST, STAR and NSTDA (Thailand); TUBITAK and TAEK (Turkey); NASU and SFFR (Ukraine); STFC (United Kingdom); DOE and NSF (USA).

Individuals have received support from the Marie-Curie programme and the European Research Council and EPLANET (European Union); the Leventis Foundation; the A. P. Sloan Foundation; the Alexander von Humboldt Foundation; the Belgian Federal Science Policy Office; the Fonds pour la Formation \`a la Recherche dans l'Industrie et dans l'Agriculture (FRIA-Belgium); the Agentschap voor Innovatie door Wetenschap en Technologie (IWT-Belgium); the Ministry of Education, Youth and Sports (MEYS) of the Czech Republic; the Council of Science and Industrial Research, India; the HOMING PLUS programme of the Foundation for Polish Science, cofinanced from European Union, Regional Development Fund; the Mobility Plus programme of the Ministry of Science and Higher Education (Poland); the OPUS programme of the National Science Center (Poland); MIUR project 20108T4XTM (Italy); the Thalis and Aristeia programmes cofinanced by EU-ESF and the Greek NSRF; the National Priorities Research Program by Qatar National Research Fund; the Programa Clar\'in-COFUND del Principado de Asturias; the Rachadapisek Sompot Fund for Postdoctoral Fellowship, Chulalongkorn University (Thailand); the Chulalongkorn Academic into Its 2nd Century Project Advancement Project (Thailand); and the Welch Foundation, contract C-1845.

\end{acknowledgments}

\bibliography{auto_generated}

\providecommand{\href}[2]{#2}\begingroup\raggedright\begin{thebibliography}{10}%
\makeatletter
\providecommand{\hrefCMSnoop }[0]{\@secondoftwo}%
\makeatother
\providecommand{\doi}{\texttt{doi:}\begingroup \urlstyle{tt}\Url}

\bibitem{Barnett:1993ea}
\hrefCMSnoop {}{R.~M. Barnett, J.~F. Gunion, and H.~E. Haber, ``{Discovering
  supersymmetry with like-sign dileptons}'',} \textit{ Phys. Lett. B} \textbf{
  315} (1993) 349,
  \href{http://dx.doi.org/10.1016/0370-2693(93)91623-U}{\doi{10.1016/0370-2693(93)91623-U}},
\href{http://www.arXiv.org/abs/hep-ph/9306204}{\texttt{arXiv:hep-ph/9306204}}.

\bibitem{Guchait:1994zk}
\hrefCMSnoop {}{M.~Guchait and D.~P. Roy, ``{Like-sign dilepton signature for
  gluino production at CERN LHC including top quark and Higgs boson
  effects}'',} \textit{ Phys. Rev. D} \textbf{ 52} (1995) 133,
  \href{http://dx.doi.org/10.1103/PhysRevD.52.133}{\doi{10.1103/PhysRevD.52.133}},
\href{http://www.arXiv.org/abs/hep-ph/9412329}{\texttt{arXiv:hep-ph/9412329}}.

\bibitem{Almeida:1997em}
F.~M.~L. Almeida, Jr.\hrefCMSnoop {}{ {et~al.}, ``{Same-sign dileptons as a
  signature for heavy Majorana neutrinos in hadron-hadron collisions}'',}
  \textit{ Phys. Lett. B} \textbf{ 400} (1997) 331,
  \href{http://dx.doi.org/10.1016/S0370-2693(97)00143-3}{\doi{10.1016/S0370-2693(97)00143-3}},
\href{http://www.arXiv.org/abs/hep-ph/9703441}{\texttt{arXiv:hep-ph/9703441}}.

\bibitem{Contino:2008hi}
\hrefCMSnoop {}{R.~Contino and G.~Servant, ``{Discovering the top partners at
  the LHC using same-sign dilepton final states}'',} \textit{ JHEP} \textbf{
  06} (2008) 026,
  \href{http://dx.doi.org/10.1088/1126-6708/2008/06/026}{\doi{10.1088/1126-6708/2008/06/026}},
\href{http://www.arXiv.org/abs/0801.1679}{\texttt{arXiv:0801.1679}}.

\bibitem{Bai:2008sk}
\hrefCMSnoop {}{Y.~Bai and Z.~Han, ``{Top-antitop and top-top resonances in the
  dilepton channel at the CERN LHC}'',} \textit{ JHEP} \textbf{ 04} (2009) 056,
  \href{http://dx.doi.org/10.1088/1126-6708/2009/04/056}{\doi{10.1088/1126-6708/2009/04/056}},
\href{http://www.arXiv.org/abs/0809.4487}{\texttt{arXiv:0809.4487}}.

\bibitem{Berger:2011ua}
E.~L. Berger\hrefCMSnoop {}{ {et~al.}, ``{Top quark forward-backward asymmetry
  and same-sign top quark pairs}'',} \textit{ Phys. Rev. Lett.} \textbf{ 106}
  (2011) 201801,
  \href{http://dx.doi.org/10.1103/PhysRevLett.106.201801}{\doi{10.1103/PhysRevLett.106.201801}},
\href{http://www.arXiv.org/abs/1101.5625}{\texttt{arXiv:1101.5625}}.

\bibitem{Ramond:1971gb}
\hrefCMSnoop {}{P.~Ramond, ``{Dual theory for free fermions}'',} \textit{ Phys.
  Rev. D} \textbf{ 3} (1971) 2415,
\href{http://dx.doi.org/10.1103/PhysRevD.3.2415}{\doi{10.1103/PhysRevD.3.2415}}.

\bibitem{Golfand:1971iw}
\href {http://www.jetpletters.ac.ru/ps/1584/article_24309.pdf}{Y.~A. Gol'fand
  and E.~P. Likhtman, ``Extension of the algebra of {P}oincar\'{e} group
  generators and violation of {P} invariance'',} \textit{ JETP Lett.} \textbf{
  13} (1971)
323.

\bibitem{Neveu:1971rx}
\hrefCMSnoop {}{A.~Neveu and J.~H. Schwarz, ``{Factorizable dual model of
  pions}'',} \textit{ Nucl. Phys. B} \textbf{ 31} (1971) 86,
\href{http://dx.doi.org/10.1016/0550-3213(71)90448-2}{\doi{10.1016/0550-3213(71)90448-2}}.

\bibitem{Volkov:1972jx}
\href {http://www.jetpletters.ac.ru/ps/1766/article_26864.pdf}{D.~V. Volkov and
  V.~P. Akulov, ``{Possible universal neutrino interaction}'',} \textit{ JETP
  Lett.} \textbf{ 16} (1972)
438.

\bibitem{Wess:1973kz}
\hrefCMSnoop {}{J.~Wess and B.~Zumino, ``{A lagrangian model invariant under
  supergauge transformations}'',} \textit{ Phys. Lett. B} \textbf{ 49} (1974)
  52,
\href{http://dx.doi.org/10.1016/0370-2693(74)90578-4}{\doi{10.1016/0370-2693(74)90578-4}}.

\bibitem{Wess:1974tw}
\hrefCMSnoop {}{J.~Wess and B.~Zumino, ``{Supergauge transformations in
  four-dimensions}'',} \textit{ Nucl. Phys. B} \textbf{ 70} (1974) 39,
\href{http://dx.doi.org/10.1016/0550-3213(74)90355-1}{\doi{10.1016/0550-3213(74)90355-1}}.

\bibitem{Fayet:1974pd}
\hrefCMSnoop {}{P.~Fayet, ``{Supergauge invariant extension of the Higgs
  mechanism and a model for the electron and its neutrino}'',} \textit{ Nucl.
  Phys. B} \textbf{ 90} (1975) 104,
\href{http://dx.doi.org/10.1016/0550-3213(75)90636-7}{\doi{10.1016/0550-3213(75)90636-7}}.

\bibitem{Nilles:1983ge}
\hrefCMSnoop {}{H.~P. Nilles, ``{Supersymmetry, supergravity and particle
  physics}'',} \textit{ Phys. Rept.} \textbf{ 110} (1984) 1,
\href{http://dx.doi.org/10.1016/0370-1573(84)90008-5}{\doi{10.1016/0370-1573(84)90008-5}}.

\bibitem{Martin:1997ns}
\hrefCMSnoop {}{S.~P. Martin, ``A supersymmetry primer'',} in \textit{
  {Perspectives on Supersymmetry II}}, G.~L. Kane, ed., p.~1.
\newblock World Scientific, 2010.
\newblock
  \href{http://www.arXiv.org/abs/hep-ph/9709356}{\texttt{arXiv:hep-ph/9709356}}.
\newblock Adv. Ser. Direct. High Energy Phys., vol. 21.
\href{http://dx.doi.org/10.1142/9789814307505_0001}{\doi{10.1142/9789814307505_0001}}.

\bibitem{ATLAS:2012ai}
\hrefCMSnoop {}{{ATLAS Collaboration}, ``{Search for gluinos in events with two
  same-sign leptons, jets and missing transverse momentum with the ATLAS
  detector in pp collisions at $\sqrt{s}=7\TeV$}'',} \textit{ Phys. Rev. Lett.}
  \textbf{ 108} (2012) 241802,
  \href{http://dx.doi.org/10.1103/PhysRevLett.108.241802}{\doi{10.1103/PhysRevLett.108.241802}},
\href{http://www.arXiv.org/abs/1203.5763}{\texttt{arXiv:1203.5763}}.

\bibitem{Aad:2014pda}
\hrefCMSnoop {}{{ATLAS Collaboration}, ``{Search for supersymmetry at
  $\sqrt{s}= 8\TeV$ in final states with jets and two same-sign leptons or
  three leptons with the ATLAS detector}'',} \textit{ JHEP} \textbf{ 06} (2014)
  035,
  \href{http://dx.doi.org/10.1007/JHEP06(2014)035}{\doi{10.1007/JHEP06(2014)035}},
\href{http://www.arXiv.org/abs/1404.2500}{\texttt{arXiv:1404.2500}}.

\bibitem{Aad:2016tuk}
\hrefCMSnoop {}{{ATLAS Collaboration}, ``{Search for supersymmetry at
  $\sqrt{s}=13$ TeV in final states with jets and two same-sign leptons or
  three leptons with the ATLAS detector}'',} \textit{ Eur. Phys. J. C} \textbf{
  76} (2016) 259,
  \href{http://dx.doi.org/10.1140/epjc/s10052-016-4095-8}{\doi{10.1140/epjc/s10052-016-4095-8}},
\href{http://www.arXiv.org/abs/1602.09058}{\texttt{arXiv:1602.09058}}.

\bibitem{SUS-10-004}
\hrefCMSnoop {}{{CMS Collaboration}, ``{Search for new physics with same-sign
  isolated dilepton events with jets and missing transverse energy at the
  LHC}'',} \textit{ JHEP} \textbf{ 06} (2011) 077,
  \href{http://dx.doi.org/10.1007/JHEP06(2011)077}{\doi{10.1007/JHEP06(2011)077}},
\href{http://www.arXiv.org/abs/1104.3168}{\texttt{arXiv:1104.3168}}.

\bibitem{SUS-11-020}
\hrefCMSnoop {}{{CMS Collaboration}, ``{Search for new physics in events with
  same-sign dileptons and b-tagged jets in pp collisions at
  $\sqrt{s}=7\TeV$}'',} \textit{ JHEP} \textbf{ 08} (2012) 110,
  \href{http://dx.doi.org/10.1007/JHEP08(2012)110}{\doi{10.1007/JHEP08(2012)110}},
\href{http://www.arXiv.org/abs/1205.3933}{\texttt{arXiv:1205.3933}}.

\bibitem{SUS-11-010}
\hrefCMSnoop {}{{CMS Collaboration}, ``{Search for new physics with same-sign
  isolated dilepton events with jets and missing transverse energy}'',}
  \textit{ Phys. Rev. Lett.} \textbf{ 109} (2012) 071803,
  \href{http://dx.doi.org/10.1103/PhysRevLett.109.071803}{\doi{10.1103/PhysRevLett.109.071803}},
\href{http://www.arXiv.org/abs/1205.6615}{\texttt{arXiv:1205.6615}}.

\bibitem{SUS-12-017}
\hrefCMSnoop {}{{CMS Collaboration}, ``{Search for new physics in events with
  same-sign dileptons and b jets in pp collisions at $\sqrt{s}=8\TeV$}'',}
  \textit{ JHEP} \textbf{ 03} (2013) 037,
  \href{http://dx.doi.org/10.1007/JHEP03(2013)037}{\doi{10.1007/JHEP03(2013)037}},
  \href{http://www.arXiv.org/abs/1212.6194}{\texttt{arXiv:1212.6194}}.
[Erratum: \DOI{10.1007/JHEP07(2013)041}].

\bibitem{SUS-13-013}
\hrefCMSnoop {}{{CMS Collaboration}, ``{Search for new physics in events with
  same-sign dileptons and jets in pp collisions at 8 TeV}'',} \textit{ JHEP}
  \textbf{ 01} (2014) 163,
  \href{http://dx.doi.org/10.1007/JHEP01(2014)163}{\doi{10.1007/JHEP01(2014)163}},
  \href{http://www.arXiv.org/abs/1311.6736}{\texttt{arXiv:1311.6736}}.

\bibitem{Farrar:1978xj}
\hrefCMSnoop {}{G.~R. Farrar and P.~Fayet, ``{Phenomenology of the production,
  decay, and detection of new hadronic states associated with
  supersymmetry}'',} \textit{ Phys. Lett. B} \textbf{ 76} (1978) 575,
\href{http://dx.doi.org/10.1016/0370-2693(78)90858-4}{\doi{10.1016/0370-2693(78)90858-4}}.

\bibitem{cmsDetector}
\hrefCMSnoop {}{{CMS Collaboration}, ``The {CMS} experiment at the {CERN}
  {LHC}'',} \textit{ JINST} \textbf{ 3} (2008) S08004,
\href{http://dx.doi.org/10.1088/1748-0221/3/08/S08004}{\doi{10.1088/1748-0221/3/08/S08004}}.

\bibitem{Chatrchyan:2012xi}
\hrefCMSnoop {}{{CMS Collaboration}, ``{Performance of CMS muon reconstruction
  in pp collision events at $\sqrt{s} = 7\TeV$}'',} \textit{ JINST} \textbf{ 7}
  (2012) P10002,
  \href{http://dx.doi.org/10.1088/1748-0221/7/10/P10002}{\doi{10.1088/1748-0221/7/10/P10002}},
\href{http://www.arXiv.org/abs/1206.4071}{\texttt{arXiv:1206.4071}}.

\bibitem{Khachatryan:2015hwa}
\hrefCMSnoop {}{{CMS Collaboration}, ``{Performance of electron reconstruction
  and selection with the CMS detector in proton-proton collisions at $\sqrt{s}
  = 8\TeV$}'',} \textit{ JINST} \textbf{ 10} (2015) P06005,
  \href{http://dx.doi.org/10.1088/1748-0221/10/06/P06005}{\doi{10.1088/1748-0221/10/06/P06005}},
\href{http://www.arXiv.org/abs/1502.02701}{\texttt{arXiv:1502.02701}}.

\bibitem{Hocker:2007ht}
\href {http://pos.sissa.it/archive/conferences/050/040/ACAT_040.pdf}{H.~Voss,
  A.~H{\"o}cker, J.~Stelzer, and F.~Tegenfeldt, ``{TMVA}, the Toolkit for
  Multivariate Data Analysis with {ROOT}'',} in \textit{ XIth International
  Workshop on Advanced Computing and Analysis Techniques in Physics Research
  (ACAT)}, p.~40.
\newblock 2007.
\newblock
\href{http://www.arXiv.org/abs/physics/0703039}{\texttt{arXiv:physics/0703039}}.
\newblock

\bibitem{Rehermann:2010vq}
\hrefCMSnoop {}{K.~Rehermann and B.~Tweedie, ``{Efficient identification of
  boosted semileptonic top quarks at the LHC}'',} \textit{ JHEP} \textbf{ 03}
  (2011) 059,
  \href{http://dx.doi.org/10.1007/JHEP03(2011)059}{\doi{10.1007/JHEP03(2011)059}},
\href{http://www.arXiv.org/abs/1007.2221}{\texttt{arXiv:1007.2221}}.

\bibitem{CMS:2009nxa}
\href {https://cds.cern.ch/record/1194487}{{CMS Collaboration},
  ``{Particle-flow event reconstruction in CMS and performance for jets, taus,
  and \MET}'',} CMS Physics Analysis Summary CMS-PAS-PFT-09-001, CERN, 2009.

\bibitem{CMS:2010byl}
\href {https://cds.cern.ch/record/1247373}{{CMS Collaboration},
  ``{Commissioning of the particle-flow event reconstruction with the first LHC
  collisions recorded in the CMS detector}'',} CMS Physics Analysis Summary
  CMS-PAS-PFT-10-001, CERN, 2010.

\bibitem{Albajar:1986iu}
\hrefCMSnoop {}{{UA1} Collaboration, ``{Beauty production at the CERN
  proton-antiproton collider}'',} \textit{ Phys. Lett. B} \textbf{ 186} (1987)
  237,
\href{http://dx.doi.org/10.1016/0370-2693(87)90287-5}{\doi{10.1016/0370-2693(87)90287-5}}.

\bibitem{JME-13-003}
\hrefCMSnoop {}{{CMS Collaboration}, ``{Performance of the CMS missing
  transverse momentum reconstruction in pp data at $\sqrt{s} = 8\TeV$}'',}
  \textit{ JINST} \textbf{ 10} (2015) P02006,
  \href{http://dx.doi.org/10.1088/1748-0221/10/02/P02006}{\doi{10.1088/1748-0221/10/02/P02006}},
\href{http://www.arXiv.org/abs/1411.0511}{\texttt{arXiv:1411.0511}}.

\bibitem{Cacciari:2008gp}
\hrefCMSnoop {}{M.~Cacciari, G.~P. Salam, and G.~Soyez, ``The anti-$k_t$ jet
  clustering algorithm'',} \textit{ JHEP} \textbf{ 04} (2008) 063,
  \href{http://dx.doi.org/10.1088/1126-6708/2008/04/063}{\doi{10.1088/1126-6708/2008/04/063}},
  \href{http://www.arXiv.org/abs/0802.1189}{\texttt{arXiv:0802.1189}}.

\bibitem{Chatrchyan:2011ds}
\hrefCMSnoop {}{{CMS Collaboration}, ``{Determination of jet energy calibration
  and transverse momentum resolution in CMS}'',} \textit{ JINST} \textbf{ 6}
  (2011) P11002,
  \href{http://dx.doi.org/10.1088/1748-0221/6/11/P11002}{\doi{10.1088/1748-0221/6/11/P11002}},
\href{http://www.arXiv.org/abs/1107.4277}{\texttt{arXiv:1107.4277}}.

\bibitem{CMS-PAS-BTV-15-001}
\href {http://cds.cern.ch/record/2138504}{{CMS Collaboration},
  ``{Identification of b quark jets at the CMS Experiment in the LHC Run 2}'',}
  CMS Physics Analysis Summary CMS-PAS-BTV-15-001, CERN, 2016.

\bibitem{MADGRAPH5}
J.~Alwall\hrefCMSnoop {}{ {et~al.}, ``{The automated computation of tree-level
  and next-to-leading order differential cross sections, and their matching to
  parton shower simulations}'',} \textit{ JHEP} \textbf{ 07} (2014) 079,
  \href{http://dx.doi.org/10.1007/JHEP07(2014)079}{\doi{10.1007/JHEP07(2014)079}},
\href{http://www.arXiv.org/abs/1405.0301}{\texttt{arXiv:1405.0301}}.

\bibitem{Melia:2011tj}
\hrefCMSnoop {}{T.~Melia, P.~Nason, R.~Rontsch, and G.~Zanderighi, ``{$\PW^+
  \PW^-$, $\PW \PZ$ and $\PZ \PZ$ production in the POWHEG BOX}'',} \textit{
  JHEP} \textbf{ 11} (2011) 078,
  \href{http://dx.doi.org/10.1007/JHEP11(2011)078}{\doi{10.1007/JHEP11(2011)078}},
\href{http://www.arXiv.org/abs/1107.5051}{\texttt{arXiv:1107.5051}}.

\bibitem{Nason:2013ydw}
\hrefCMSnoop {}{P.~Nason and G.~Zanderighi, ``{$\PW^+ \PW^-$ , $\PW \PZ$ and
  $\PZ \PZ$ production in the POWHEG-BOX-V2}'',} \textit{ Eur. Phys. J. C}
  \textbf{ 74} (2014) 2702,
  \href{http://dx.doi.org/10.1140/epjc/s10052-013-2702-5}{\doi{10.1140/epjc/s10052-013-2702-5}},
\href{http://www.arXiv.org/abs/1311.1365}{\texttt{arXiv:1311.1365}}.

\bibitem{Ball:2014uwa}
\hrefCMSnoop {}{{NNPDF} Collaboration, ``{Parton distributions for the LHC Run
  II}'',} \textit{ JHEP} \textbf{ 04} (2015) 040,
  \href{http://dx.doi.org/10.1007/JHEP04(2015)040}{\doi{10.1007/JHEP04(2015)040}},
\href{http://www.arXiv.org/abs/1410.8849}{\texttt{arXiv:1410.8849}}.

\bibitem{Sjostrand:2007gs}
\hrefCMSnoop {}{T.~Sj{\"o}strand, S.~Mrenna, and P.~Z. Skands, ``{A brief
  introduction to PYTHIA 8.1}'',} \textit{ Comput. Phys. Commun.} \textbf{ 178}
  (2008) 852,
  \href{http://dx.doi.org/10.1016/j.cpc.2008.01.036}{\doi{10.1016/j.cpc.2008.01.036}},
\href{http://www.arXiv.org/abs/0710.3820}{\texttt{arXiv:0710.3820}}.

\bibitem{Skands:2014pea}
\hrefCMSnoop {}{P.~Skands, S.~Carrazza, and J.~Rojo, ``{Tuning PYTHIA 8.1: the
  Monash 2013 tune}'',} \textit{ Eur. Phys. J. C} \textbf{ 74} (2014) 3024,
  \href{http://dx.doi.org/10.1140/epjc/s10052-014-3024-y}{\doi{10.1140/epjc/s10052-014-3024-y}},
\href{http://www.arXiv.org/abs/1404.5630}{\texttt{arXiv:1404.5630}}.

\bibitem{CMS-PAS-GEN-14-001}
\hrefCMSnoop {}{{CMS Collaboration}, ``{Event generator tunes obtained from
  underlying event and multiparton scattering measurements}'',} \textit{ Eur.
  Phys. J. C} \textbf{ 76} (2016) 155,
  \href{http://dx.doi.org/10.1140/epjc/s10052-016-3988-x}{\doi{10.1140/epjc/s10052-016-3988-x}},
\href{http://www.arXiv.org/abs/1512.00815}{\texttt{arXiv:1512.00815}}.

\bibitem{Geant}
\hrefCMSnoop {}{{GEANT4} Collaboration, ``{GEANT4---a simulation toolkit}'',}
  \textit{ Nucl. Instrum. Meth. A} \textbf{ 506} (2003) 250,
\href{http://dx.doi.org/10.1016/S0168-9002(03)01368-8}{\doi{10.1016/S0168-9002(03)01368-8}}.

\bibitem{Abdullin:2011zz}
S.~Abdullin\hrefCMSnoop {}{ {et~al.}, ``{The fast simulation of the CMS
  detector at LHC}'',} \textit{ J. Phys. Conf. Ser.} \textbf{ 331} (2011)
  032049,
\href{http://dx.doi.org/10.1088/1742-6596/331/3/032049}{\doi{10.1088/1742-6596/331/3/032049}}.

\bibitem{Alves:2011wf}
\hrefCMSnoop {}{D.~Alves {et~al.}, ``{Simplified models for LHC new physics
  searches}'',} \textit{ J. Phys. G} \textbf{ 39} (2012) 105005,
  \href{http://dx.doi.org/10.1088/0954-3899/39/10/105005}{\doi{10.1088/0954-3899/39/10/105005}},
\href{http://www.arXiv.org/abs/1105.2838}{\texttt{arXiv:1105.2838}}.

\bibitem{Chatrchyan:2013sza}
\hrefCMSnoop {}{{CMS Collaboration}, ``{Interpretation of searches for
  supersymmetry with simplified models}'',} \textit{ Phys. Rev. D} \textbf{ 88}
  (2013) 052017,
  \href{http://dx.doi.org/10.1103/PhysRevD.88.052017}{\doi{10.1103/PhysRevD.88.052017}},
\href{http://www.arXiv.org/abs/1301.2175}{\texttt{arXiv:1301.2175}}.

\bibitem{Arnison:1983rp}
\hrefCMSnoop {}{{UA1} Collaboration, ``{Experimental observation of isolated
  large transverse energy electrons with associated missing energy at $\sqrt{s}
  = 540\GeV$}'',} \textit{ Phys. Lett. B} \textbf{ 122} (1983) 103,
\href{http://dx.doi.org/10.1016/0370-2693(83)91177-2}{\doi{10.1016/0370-2693(83)91177-2}}.

\bibitem{ATLAS:2014kca}
\hrefCMSnoop {}{{ATLAS Collaboration}, ``{Search for anomalous production of
  prompt same-sign lepton pairs and pair-produced doubly charged Higgs bosons
  with $ \sqrt{s}=8\TeV$ pp collisions using the ATLAS detector}'',} \textit{
  JHEP} \textbf{ 03} (2015) 041,
  \href{http://dx.doi.org/10.1007/JHEP03(2015)041}{\doi{10.1007/JHEP03(2015)041}},
\href{http://www.arXiv.org/abs/1412.0237}{\texttt{arXiv:1412.0237}}.

\bibitem{CMS-PAS-LUM-15-001}
\href {https://cds.cern.ch/record/2138682}{{CMS Collaboration}, ``{CMS
  luminosity measurement for the 2015 data taking period}'',} CMS Physics
  Analysis Summary CMS-PAS-LUM-15-001, CERN, 2016.

\bibitem{Read:2002hq}
\hrefCMSnoop {}{A.~L. Read, ``Presentation of search results: the {$CL_s$}
  technique'',} \textit{ J. Phys. G} \textbf{ 28} (2002) 2693,
\href{http://dx.doi.org/10.1088/0954-3899/28/10/313}{\doi{10.1088/0954-3899/28/10/313}}.

\bibitem{Junk:1999kv}
\hrefCMSnoop {}{T.~Junk, ``{Confidence level computation for combining searches
  with small statistics}'',} \textit{ Nucl. Instrum. Meth. A} \textbf{ 434}
  (1999) 435,
  \href{http://dx.doi.org/10.1016/S0168-9002(99)00498-2}{\doi{10.1016/S0168-9002(99)00498-2}},
\href{http://www.arXiv.org/abs/hep-ex/9902006}{\texttt{arXiv:hep-ex/9902006}}.

\bibitem{ATL-PHYS-PUB-2011-011}
\href {https://cdsweb.cern.ch/record/1379837}{{ATLAS and CMS Collaborations},
  ``Procedure for the LHC Higgs boson search combination in summer 2011'',}
  Technical Report {CMS NOTE-2011/005}, CERN, 2011.

\bibitem{Cowan:2010js}
\hrefCMSnoop {}{G.~Cowan, K.~Cranmer, E.~Gross, and O.~Vitells, ``{Asymptotic
  formulae for likelihood-based tests of new physics}'',} \textit{ Eur. Phys.
  J. C} \textbf{ 71} (2011) 1554,
  \href{http://dx.doi.org/10.1140/epjc/s10052-011-1554-0}{\doi{10.1140/epjc/s10052-011-1554-0}},
  \href{http://www.arXiv.org/abs/1007.1727}{\texttt{arXiv:1007.1727}}.
[Erratum: \DOI{10.1140/epjc/s10052-013-2501-z}].

\bibitem{bib-nlo-nll-01}
\hrefCMSnoop {}{W.~Beenakker, R.~H{\"o}pker, M.~Spira, and P.~M. Zerwas,
  ``Squark and gluino production at hadron colliders'',} \textit{ Nucl. Phys.
  B} \textbf{ 492} (1997) 51,
  \href{http://dx.doi.org/10.1016/S0550-3213(97)80027-2}{\doi{10.1016/S0550-3213(97)80027-2}},
\href{http://www.arXiv.org/abs/hep-ph/9610490}{\texttt{arXiv:hep-ph/9610490}}.

\bibitem{bib-nlo-nll-02}
\hrefCMSnoop {}{A.~Kulesza and L.~Motyka, ``{Threshold resummation for
  squark-antisquark and gluino-pair production at the LHC}'',} \textit{ Phys.
  Rev. Lett.} \textbf{ 102} (2009) 111802,
  \href{http://dx.doi.org/10.1103/PhysRevLett.102.111802}{\doi{10.1103/PhysRevLett.102.111802}},
\href{http://www.arXiv.org/abs/0807.2405}{\texttt{arXiv:0807.2405}}.

\bibitem{bib-nlo-nll-03}
\hrefCMSnoop {}{A.~Kulesza and L.~Motyka, ``{Soft gluon resummation for the
  production of gluino-gluino and squark-antisquark pairs at the {LHC}}'',}
  \textit{ Phys. Rev. D} \textbf{ 80} (2009) 095004,
  \href{http://dx.doi.org/10.1103/PhysRevD.80.095004}{\doi{10.1103/PhysRevD.80.095004}},
\href{http://www.arXiv.org/abs/0905.4749}{\texttt{arXiv:0905.4749}}.

\bibitem{bib-nlo-nll-04}
W.~Beenakker\hrefCMSnoop {}{ {et~al.}, ``{Soft-gluon resummation for squark and
  gluino hadroproduction}'',} \textit{ JHEP} \textbf{ 12} (2009) 041,
  \href{http://dx.doi.org/10.1088/1126-6708/2009/12/041}{\doi{10.1088/1126-6708/2009/12/041}},
\href{http://www.arXiv.org/abs/0909.4418}{\texttt{arXiv:0909.4418}}.

\bibitem{bib-nlo-nll-05}
W.~Beenakker\hrefCMSnoop {}{ {et~al.}, ``{Squark and gluino
  hadroproduction}'',} \textit{ Int. J. Mod. Phys. A} \textbf{ 26} (2011) 2637,
  \href{http://dx.doi.org/10.1142/S0217751X11053560}{\doi{10.1142/S0217751X11053560}},
\href{http://www.arXiv.org/abs/1105.1110}{\texttt{arXiv:1105.1110}}.

\bibitem{Borschensky:2014cia}
C.~Borschensky\hrefCMSnoop {}{ {et~al.}, ``{Squark and gluino production cross
  sections in pp collisions at $\sqrt{s} =$ 13, 14, 33 and 100\TeV}'',}
  \textit{ Eur. Phys. J. C} \textbf{ 74} (2014) 3174,
  \href{http://dx.doi.org/10.1140/epjc/s10052-014-3174-y}{\doi{10.1140/epjc/s10052-014-3174-y}},
\href{http://www.arXiv.org/abs/1407.5066}{\texttt{arXiv:1407.5066}}.

\end{thebibliography}\endgroup

\cleardoublepage \appendix\section{The CMS Collaboration \label{app:collab}}\begin{sloppypar}\hyphenpenalty=5000\widowpenalty=500\clubpenalty=5000\textbf{Yerevan Physics Institute,  Yerevan,  Armenia}\\*[0pt]
V.~Khachatryan, A.M.~Sirunyan, A.~Tumasyan
\vskip\cmsinstskip
\textbf{Institut f\"{u}r Hochenergiephysik der OeAW,  Wien,  Austria}\\*[0pt]
W.~Adam, E.~Asilar, T.~Bergauer, J.~Brandstetter, E.~Brondolin, M.~Dragicevic, J.~Er\"{o}, M.~Flechl, M.~Friedl, R.~Fr\"{u}hwirth\cmsAuthorMark{1}, V.M.~Ghete, C.~Hartl, N.~H\"{o}rmann, J.~Hrubec, M.~Jeitler\cmsAuthorMark{1}, A.~K\"{o}nig, I.~Kr\"{a}tschmer, D.~Liko, T.~Matsushita, I.~Mikulec, D.~Rabady, N.~Rad, B.~Rahbaran, H.~Rohringer, J.~Schieck\cmsAuthorMark{1}, J.~Strauss, W.~Treberer-Treberspurg, W.~Waltenberger, C.-E.~Wulz\cmsAuthorMark{1}
% ...
\vskip\cmsinstskip
% ...
1:~~Also at Vienna University of Technology, Vienna, Austria\\
% ...

\end{sloppypar}
\end{document}